\documentclass[aps,preprintnumbers,onecolumn,amsmath,amssymb,floatfix,pra]{revtex4} 

\pdfoutput=1

\usepackage{graphicx}
\usepackage{color}
\usepackage{mathrsfs}
\usepackage{amsmath}
\usepackage{braket}
\usepackage{verbatim}
\usepackage{enumitem}
\usepackage{ifpdf}

\ifpdf
\usepackage[pdftex,breaklinks=true,colorlinks,urlcolor=blue,citecolor=blue]{hyperref}
\else
\usepackage[hypertex,ps2pdf,dvips,colorlinks,urlcolor=blue,citecolor=blue]{hyperref}
\fi
\newcommand{\e}{\mathrm e}
\renewcommand{\i}{\mathrm i}
\renewcommand{\a}{\alpha}
\newcommand{\cO}{\mathcal{O}}
\newcommand\epc{\,,}
\newcommand\epp{\,.}

\begin{document}

\title{Overlap distributions for quantum quenches in the anisotropic Heisenberg chain}

\author{
Paolo P. Mazza,$^1$ 
Jean-Marie St\'ephan,$^{1}$ 
Elena Canovi,$^2$
Vincenzo Alba,$^{3}$
Michael Brockmann,$^1$
and Masudul Haque$^1$}

\affiliation{
$^1$Max Planck Institute for the Physics of Complex Systems, N{\"o}thnitzer Str.~38, 01187 Dresden, Germany\\
$^2$Max Planck Research Department for Structural Dynamics, University of Hamburg-CFEL, Hamburg, Germany\\
$^3$SISSA, via Bonomea 265, 34136 Trieste, Italy, INFN, sezione di Trieste
}

\date{\today}

\begin{abstract}
  The dynamics after a quantum quench is determined by the weights of the initial state in the
  eigenspectrum of the final Hamiltonian, i.e., by the distribution of overlaps in the energy
  spectrum.  We present an analysis of such overlap distributions for quenches of the anisotropy
  parameter in the one-dimensional anisotropic spin-1/2 Heisenberg model (XXZ chain).  We provide an
  overview of the form of the overlap distribution for quenches from various initial anisotropies to
  various final ones, using numerical exact diagonalization.  We show that if the system is
  prepared in the antiferromagnetic N\'eel state (infinite anisotropy) and released into a
  non-interacting setup (zero anisotropy, XX point) only a small fraction of the final eigenstates gives
  contributions to the post-quench dynamics, and that these eigenstates have identical overlap
  magnitudes.  We derive expressions for the overlaps, and present the selection rules that determine
  the final eigenstates having nonzero overlap.  We use these results to derive concise
  expressions for time-dependent quantities (Loschmidt echo, longitudinal and transverse
  correlators) after the quench.  We use perturbative analyses to understand the overlap
  distribution for quenches from infinite to small nonzero anisotropies, and for quenches from large
  to zero anisotropy.
\end{abstract}


\maketitle

\section{Introduction and motivation}

The past decade has witnessed a tremendous resurgence of interest in the physics of quantum
many-body systems out of equilibrium \cite{
  2011_Polkovnikov_RevModPhys_83_863,
  2015_Eisert_NatPhys_11_124, 
  2010_Dziarmaga_AdvPhys_59_1063}.  
This growth has been partly motivated
by remarkable developments in experiments, especially in experiments on cold atoms, which made the
explicit observation of time evolution of many-body quantum systems possible \cite{
  2002_Greiner_Nature_419_51, 2006_Kinoshita_Nature_440_900, 2012_Schneider_NaturePhys_8_213,
  2012_Trotzky_Nature_8_325, 2007_Hofferberth_Nature_449_324, 2012_Cheneau_Nature_481_484,
  2012_Gring_Science_337_1318, 2013_Ronzheimer_PRL_110_205301, 2007_Lee_PRL_99_020402,
  2009_Palzer_PRL_103_150601, 2007_Trotzky_Science_319_295}.  
In addition to the
experimental motivation, fundamental conceptual issues, such as thermalization in isolated systems
and the nature of adiabaticity, have also played a strong role in driving the development of this
field.  A central paradigm in the study of non-equilibrium physics is the \emph{quantum quench}.  A quantum
quench involves a protocol starting from the ground state (or a thermal state or another eigenstate) of an
`initial' Hamiltonian $H_i$, and then rapidly changing the Hamiltonian so that nontrivial time
evolution occurs under a different (`final') Hamiltonian $H_f$.  Many aspects of the dynamics
induced by quantum quenches have been considered in recent years; the literature is rapidly growing
and already too vast to review here.

A quantum quench can lead to nontrivial dynamics as the initial state generally has overlap
with many eigenstates of the final Hamiltonian.  If $|n\rangle$ and $E_n$ are the eigenstates and
eigenvalues of the final Hamiltonian, i.e., $H_f|n\rangle = E_n|n\rangle$, an initial state
$|\Psi_0\rangle$ will lead to the following time dependence of the system wavefunction:
\begin{equation}
  |\Psi(t)\rangle = \e^{-\i H_{f} t}|\Psi_0\rangle = \e^{-\i H_{f} t}\sum_n c_n |n\rangle = \sum_n \e^{-\i
    E_n t}c_n|n\rangle \qquad\textnormal{with}\quad c_n=\langle n|\Psi_0\rangle \epp 
\end{equation}
If a large number of final eigenstates have non-negligible overlap $|c_n|$, this sum can lead to
highly nontrivial time dependence.  Clearly, both the overlaps $|c_n| = |\langle\Psi_0|n\rangle|$
and the corresponding eigenenergies $E_n$ are important in determining the subsequent dynamics.  The
object of study of the present paper is the distribution of overlaps, i.e., the overlap magnitudes
$|c_n|$ as a function of energy $E_n$.  We will refer to this as the ``overlap distribution''.  Note
that, if the squares $|c_n|^2$ were considered, this could be regarded as a discrete probability
distribution, as $\sum_n|c_n|^2=1$ and the squares $|c_n|^2$ represent probabilities.  Since $|c_n|$
and $|c_n|^2$ provide equivalent information, we will focus on $|c_n|$.  Also note that the
distribution of phases of $c_n$ may provide useful information but are not individually
well-defined, since the phases of individual eigenstates $|n\rangle$ can be chosen
arbitrarily. 

This work presents an investigation of the features of the overlap distribution for various quenches in
a particular system.  In this introductory section, we provide motivation for the study of the
overlap distribution, by reviewing its connections to topics that have attracted widespread
attention in recent years.

The overlap distribution is closely related to the statistics of the work performed in a quench,
which is characterized by the so-called work distribution \cite{
  1997_Jarzynski_PRL_78_2690,
  2007_Talkner_PRE_75_050102, 
  2008_Silva_PRL_101_120603, 
  2011_Campisi_RMP_83_771}:
\begin{equation}
  p(w) = 2\pi \sum_n |c_n|^2 \delta(w-(E_n-E_{i}))\epc
\end{equation}
where $E_i$ is the initial energy.  The work distribution is thus simply a continuous version of the
distribution of squared overlaps, with the zero of the energy shifted by $E_i$.  The work statistics in
non-equilibrium processes in various quantum many-body systems is a topic of much recent interest
\cite{
  2008_Silva_PRL_101_120603, 2014_Shchadilova_PRL_112_070601, 2012_Dora_PRB_86_161109,
  2009_Paraan_PRE_80_061130, 2010_Mossel_NewJPhys_12_055028, 2011_Gambassi_1106.2671,
  2012_Heyl_PRL_108_190601, 2012_Gambassi_PRL_109_250602, 2012_Heyl_PRB_85_155413,
  2012_Smacchia_PRL_109_037202, 2013_Dorner_PRL_110_230601, 2013_Mazzola_PRL_110_230602,
  2013_Sotiriadis_PRE_87_052129, 2013_Smacchia_PRE_88_042109, 2014_Sindona_NJP_16_045013,
  2014_Marino_PRB_89_024303, 2014_Palmai_PRE_90_052102, 2014_Fusco_PRX_4_031029,
  2015_Dutta_PRE_92_012104, 2015_Palmai_1506.08200}.
Another quantity of wide current interest, closely related to the overlap distribution and work
distribution, is the return probability, 
\begin{equation}
  \mathcal{L}(t) = |\langle\Psi_0|\e^{-\i H_f t}|\Psi_0\rangle|^2  = |\langle\Psi_0|\e^{\i H_i t}\e^{-\i H_f t}|\Psi_0\rangle|^2 \epp
\label{eq:LoschEch}
\end{equation}
This is a special case of the Loschmidt echo where the reverse time evolution occurs under the
Hamiltonian $H_i$, for which the initial state $|\Psi_0\rangle$ is an eigenstate.  Following common
practice, we will refer to this quantity simply as the Loschmidt echo.  The behavior of the
Loschmidt echo after a quantum quench has been studied in a large body of recent literature, see,
e.g.,  
\cite{
  2008_Silva_PRL_101_120603, 2014_Andraschko_PRB_89_125120, 2012_Heyl_PRB_85_155413,
  2014_Palmai_PRE_90_052102, 2011_Stephan_JStatMech_P08019, 2010_Mossel_NewJPhys_12_055028,
  2013_Pozsgay_JStatMech_P10028, 2006_Quan_PRL_96_140604, 2014_Torres-Herrera_NJP_16_063010,
  2015_Torres-Herrera_1506.08904, 2015_Torres-Herrera_PRB_92_014208,
  2014_Torres-Herrera_PRA_90_033623, 2014_Torres-Herrera_PRE_89_062110,
  2014_Torres-Herrera_PRA_89_043620, 2013_Torres-Herrera_PRE_88_042121, 2015_Viti_1507.08132,
  2014_DeLuca_90_081403, 2013_Heyl_PRL_110_135704, 2014_Kennes_PRB_90_115101,
  2014_Canovi_PRL_113_265702, 2015_Schmitt_PRB_92_075114, 2015_Rajak_1507.00963,
  2012_Mukherjee_PRB_86_020301, 2012_Montes_PRE_86_021101, 2011_Jacobson_PRA_84_022115,
  2011_CamposVenuti_PRL_107_010403, 2012_Happola_PRA_85_032114, 2010_CamposVenuti_PRA_81_022113,
  2014_Campbell_PRA_90_013617, 2014_Vasseur_PRX_4_041007, 2013_Heyl_1310.6226,
  2014_Sachdeva_PRB_90_045421, 2013_Karrasch_87_195104, 2013_Stephan_JSTAT_P09002,
  2012_Lelas_PRA_86_033620, 2014_Kriel_PRB_90_125106, 2014_Heyl_PRL_113_205701,
  2014_Hickey_PRB_89_054301, 2013_Dora_PRL_111_046402, 2014_Dora_1410.5954, 2013_Fagotti_1308.0277,
  2014_Vajna_PRB_89_161105, 2014_Vajna_PRB_91_155127}.  
In particular, the Loschmidt echo is central to the study of so-called dynamical phase transitions \cite{
  2013_Heyl_PRL_110_135704,
  2014_Canovi_PRL_113_265702, 2013_Karrasch_87_195104, 2014_Hickey_PRB_89_054301,
  2014_Heyl_PRL_113_205701, 2014_Andraschko_PRB_89_125120, 2015_Schmitt_PRB_92_075114,
  2014_Kriel_PRB_90_125106, 2014_Vajna_PRB_89_161105, 2014_Vajna_PRB_91_155127}.  
By expanding $|\Psi_0\rangle$ in terms of the eigenstates $|n\rangle$ of the final Hamiltonian $H_f$, we find
\begin{equation}
  \mathcal{L}(t) = \left| \sum_{n}\textnormal{e}^{-\i E_n t}|c_n|^2 \right|^2 
  = \left|\int{}\frac{d\omega{}}{2\pi}p(\omega)\e^{-\i{\omega}t}\right|^2  \epp
\label{eq:LoschEch1}
\end{equation}
Knowledge of the overlap distribution, i.e., the $|c_n|$'s and corresponding $E_n$'s, thus leads
directly to the Loschmidt echo.  For example, the energy width of the distribution determines the
coefficient of the initial quadratic decay of $\mathcal{L}(t)$, as it can be seen by expanding 
$\e^{-\i H_f t}$ in Eq.~\eqref{eq:LoschEch} to second order in $t$ \cite{2015_Viti_1507.08132,2014_Torres-Herrera_PRA_89_043620}.

After a quench, a generic observable  $\cO$ has the following time evolution: 
\begin{equation}
  \langle \cO(t)\rangle = \langle\Psi(t)|\cO|\Psi(t)\rangle = \langle\Psi_0|\e^{\i H_f t} \cO \e^{-\i H_f t}|\Psi_0\rangle =
  \sum_{m,n}c_m^{\ast}c_n \e^{-\i(E_n-E_m)t}\langle m|\cO|n\rangle
\xrightarrow{t\to\infty}\sum_{n}|c_n|^2 \langle n|\cO|n\rangle \epp 
\label{eq:observable_longtime}
\end{equation}
The long-time behavior of such expectation values, $\lim_{t\to\infty}\langle\cO(t)\rangle$, is of
fundamental interest since this is closely related to the question of whether an isolated quantum
system thermalizes in some sense.  As outlined above, this long-time limit is determined solely by
the overlap magnitudes and the diagonal matrix elements.  This is often called the ``diagonal
ensemble'' value of the long-time limit \cite{2008_Rigol_Nature_452_854}. The eigenstate
thermalization hypothesis (ETH), which is expected to hold for non-integrable systems, postulates that the
diagonal matrix elements $\langle n|\cO|n\rangle $ are smooth functions of eigenenergy, for large
enough systems \cite{
  1994_Srednicki_PRE_50_888, 1991_Deutsch_PRA_43_2046, 2008_Rigol_Nature_452_854,
  2012_Neuenhahn_PRE_85_060101, 2014_Beugeling_PRE_89_042112}.  
The question of thermalization can
then reduce to the question of how narrow in energy the overlap distribution is. Recent work (such as the ``quench action'' approach \cite{
  2013_Caux_PRL_110_257203,
  2014_DeNardis_PRA_89_033601, 2014_Wouters_PRL_113_117202, 2014_Pozsgay_PRL_113_117203,
  2014_Brockmann_JStatMech_P12009, 2015_DeLuca_PRA_91_021603, 2015_Ilievski_1507.02993} 
and an approach based on linked-cluster expansions \cite{2014_Rigol_PRL_112_170601, 2014_Rigol_PRE_90_031301}) 
has formulated calculations of long-time expectation values directly in
the thermodynamic limit, thus avoiding the explicit calculation of overlaps in finite-size systems.
The diagonal ensemble formulation of Eq.~\eqref{eq:observable_longtime} for finite sizes, involving
the overlap distribution $|c_n|$, will play an essential role in the discussion of such calculations
for long-time stationary state values of observables, in particular for comparing the ordering of
the limiting procedures (limits of large sizes and large times).  We also note that the overlaps
$c_n$ appear not only in the long-time limit, but also in the arbitrary-time expression for
$\langle\cO(t)\rangle$ (cf. Eq.~\eqref{eq:observable_longtime}), although in this case the phases of 
$c_n$ are also important.

From the discussion above, it is clear that the overlap distribution plays a central role in many
aspects of quench dynamics.  Numerical data on overlap distributions have appeared in many different
non-equilibrium studies for specific quenches, e.g., in Refs.~\cite{
  2015_Torres-Herrera_1506.08904,
  2015_Torres-Herrera_PRB_92_014208, 2014_Torres-Herrera_PRA_90_033623,
  2014_Torres-Herrera_PRE_89_062110, 2013_Torres-Herrera_PRE_88_042121,
  2014_Torres-Herrera_NJP_16_063010, 2014_Torres-Herrera_PRA_89_043620} 
where the overlap distributions are called ``local density of states'', in Refs.~\cite{
  2008_Rigol_Nature_452_854,
  2011_Cassidy_PRL_106_140405, 2011_Rigol_PRA_84_033640, 2011_Santos_PRL_107_040601,
  2014_Sorg_PRA_90_033606, 2009_Rigol_PRA_80_053607, 2010_Biroli_PRL_105_250401} 
in the context of thermalization, in Ref.~\cite{2014_Shchadilova_PRL_112_070601} in connection with intermediate-time
behaviors of the Loschmidt echo, and in Ref.~\cite{2014_Hild_PRL_113_147205} for a spin spiral
initial state.  There have been, however, relatively few systematic studies of the overlap
distribution so far.  The present work is a step towards addressing this gap in the literature.

We provide a detailed study of the overlap distribution for a well-known model system, namely the
one-dimensional anisotropic spin-1/2 Heisenberg model (XXZ chain) at zero magnetization.  We
consider quenches of the anisotropy parameter $\Delta$, see Eq.~\eqref{eq:ham}, and provide an
overview of the overlap distributions obtained in quenches from various initial $\Delta_i$ values to
various final $\Delta_f$ values, using full numerical diagonalization.  For the special case of
$\Delta_i=\infty$ and $\Delta_f=0$, i.e., quenching from the N\'eel state to the so-called XX point,
we present a full analytical study of the overlap distribution. The XX point is mappable to a chain of free fermions,
which allows this detailed analysis.  We find that all nonzero overlaps are exactly equal and derive
expressions for them, using the Slater determinant structure of the XX eigenfunctions, as well as
using the Bethe ansatz.  We derive the selection rules determining the eigenstates that have nonzero
overlaps for this quench, both in the language of fermionic momenta and using the language of Bethe
roots.  We show how these results on the overlap distribution can be used to derive explicit
expressions for the Loschmidt echo $\mathcal{L}(t)$ and for equal-time correlators after the quench.
For quenches ending at small but nonzero anisotropy, $\Delta_f\ll1$, a splitting structure appears
in the overlap distribution; we analyze the main features of this structure using perturbation
theory. Similarly, for quenches to the XX point starting from large but finite initial value
$\Delta_i \gg 1$, a perturbative expansion in $1/\Delta_i$ explains the structure of the overlap
distribution. 

After introducing the model and some notation in Sec.~\ref{sec:model}, we provide our numerical
overview of overlap distributions in quenches from various $\Delta_i$ to various $\Delta_f$ in
Sec.~\ref{sec:Numerical}.  Section \ref{sec:B} is devoted to a detailed analysis of the
$\Delta_i=\infty \to \Delta_f=0$ case.  We also use the results for the overlap distribution to
derive analytical expressions for time-dependent quantities, in particular, for the Loschmidt echo and
for longitudinal and transverse two-point correlators. Section \ref{sec:A} outlines a perturbative 
treatment of the $\Delta_i=\infty \to \Delta_f\ll 1$
case (perturbation in $\Delta_f$), while Sec.~\ref{sec:C} describes a perturbative treatment of the
$\Delta_i\gg1 \to \Delta_f=0$ case (perturbation in $1/\Delta_i$).

\section{The model}\label{sec:model}

This work focuses on the one-dimensional anisotropic spin-1/2 Heisenberg model (XXZ
chain) with nearest-neighbor interactions.  The Hamiltonian is given by
\begin{equation}
  H = \sum_{j=1}^L\left(S_j^xS_{j+1}^x + S_j^yS_{j+1}^y + \Delta S_{j}^zS_{j+1}^z\right) 
  = \frac{1}{2}\sum_{j=1}^L\left(S_j^+S_{j+1}^- + S_j^-S_{j+1}^+ \right) + \Delta \sum_{j=1}^L S_{j}^zS_{j+1}^z
  = H_{\text{xx}} + \Delta H_z \epc
\label{eq:ham}
\end{equation}
where $S_j^\alpha=\sigma_j^\alpha/2$, $\alpha=x,y,z$, with $\sigma_j^\alpha$ being the usual Pauli
matrices, and $S_j^{\pm} = S_j^x \pm \i S_j^y$.  We will refer to the $\Delta=0$ part of the Hamiltonian
as the XX Hamiltonian: $H_{\text{xx}}= \frac{1}{2}\sum_{j=1}^L\left(S_j^+S_{j+1}^- + S_j^-S_{j+1}^+ \right)$.
Throughout this manuscript we choose the number $L$ of lattice sites to be even. We also distinguish
between two different types of boundary conditions. For an open chain we set $S_{L+1}^{\alpha}=0$,
for a periodic chain $S_{L+1}^{\alpha}=S_{1}^{\alpha}$, $\alpha = x,y,z$.

The XXZ chain is a natural generalization of the isotropic Heisenberg
chain~\cite{1928_Heisenberg_ZPhys_49_619}.  It is one of the most widely studied models in quantum
magnetism in particular, and condensed matter physics in general.  It is absolutely central to the
study of integrability, having a relatively simple integrable structure, yet possessing rich physics
and displaying a phase transition.  In the current era of non-equilibrium physics, it has been
widely used as a model system to explore and exemplify non-equilibrium phenomena, including studies
of the Loschmidt echo (e.g.~\cite{
  2010_Mossel_NewJPhys_12_055028, 2014_Andraschko_PRB_89_125120,
  2013_Dora_PRL_111_046402, 2014_Dora_1410.5954, 2013_Fagotti_1308.0277, 2014_Heyl_PRL_113_205701,
  2014_Torres-Herrera_PRA_89_043620}), 
studies of quenches of the anisotropy parameter $\Delta$ (e.g.~\cite{
  2010_Barmettler_NewJPhys_12_055017, 2015_Collura_1507.03492}), 
and of quenches starting from a N\'eel state (e.g.~\cite{
  2014_Wouters_PRL_113_117202, 2010_Barmettler_NewJPhys_12_055017, 2014_Fagotti_PRB_89_125101,
  2014_Pozsgay_PRL_113_117203, 2014_Brockmann_JStatMech_P12009, 2014_Heyl_PRL_113_205701,
  2014_Torres-Herrera_PRA_89_043620}).  
Dynamics starting from various non-uniform states (such as domain wall initial states), 
leading to propagation or transmission along the XXZ chain, is also a rapidly growing 
topic of investigation (see, e.g., \cite{
  2014_Liu_PRL_112_257204,
  2014_Alba_PRB_90_075144, 2005_Gobert_PRE_71_036102, 2010_Mossel_NewJPhys_12_055028,
  2013_Misguich_PRB_88_245114, 2013_Ganahl_1302.2667, 2012_Woellert_PRB_85184433,
  2009_Langer_PRB_79_214409, 2010_Haque_PRA_82_012108, 2011_Jesenko_PRB_84_174438,
  2014_Karrasch_PRB_89_075139, 2014_Sharma_PRA_89_043608, 2014_DeLuca_PRB_90_161101,
  2013_Karrasch_PRB_88_195129, 2012_Karrasch_PRL_108_227206, 2011_Langer_PRB_84_205115, 2015_Vlijm_1507.08624}). 
The XXZ chain is also being increasingly used to study real-time thermal transport and real-time dynamics at
finite temperatures \cite{
  2012_Karrasch_PRL_108_227206, 2014_Karrasch_PRB_89_075139,
  2013_Karrasch_PRB_88_195129, 2014_Bonnes_PRL_113_187203, 2011_Langer_PRB_84_205115, 
  2014_DeLuca_PRB_90_161101}.  
It is a primary testbed for developments and applications of
Bethe ansatz techniques for addressing non-equilibrium issues (e.g.~\cite{
  2010_Mossel_NewJPhys_12_055028, 2014_Goldstein_90_043625, 2014_Liu_PRL_112_257204,
  2015_Alba_91_155123, 2014_Pozsgay_JStatMech_P06011, 2014_Brockmann_JPA_47_145003,
  2014_Brockmann_JPA_47_345003, 2013_Caux_PRL_110_257203, 2014_DeNardis_PRA_89_033601,
  2014_Wouters_PRL_113_117202, 2015_Vlijm_1507.08624, 2014_Pozsgay_PRL_113_117203,
  2014_Brockmann_JStatMech_P12009, 2013_Alba_2013_JStatMech_P10018}).
We therefore find the XXZ chain to be an appropriate model for a systematic study of the work
distribution in quantum quenches. 

In this work, we restrict to positive anisotropies $\Delta>0$.  The isotropic critical point $\Delta=1$
separates two regimes with different behaviors; in terms of low-energy physics, the $\Delta<1$
regime is gapless, while the $\Delta>1$ regime is gapped.  The Hamiltonian \eqref{eq:ham} commutes
with the $z$-component $S^z=\sum_{j=1}^L S_j^z$ of the total spin operator.  For even $L$ the zero
magnetization sector contains the ground state for all $\Delta>1$.  Our initial states are ground
states of the XXZ chain and, hence, lie in the zero magnetization sector.  Thus, we can restrict to
the subspace spanned by energy eigenstates $|n\rangle$ with $S^z|n\rangle=0$.  At $\Delta=0$ the
chain can be mapped onto a gas of free fermions via the Jordan-Wigner mapping, and subsequently
solved using free-fermion techniques.  For generic values of the anisotropy parameter $\Delta$ the
XXZ model is integrable and can be solved by means of Bethe Ansatz
techniques~\cite{1931_Bethe_ZPhys_71_205, BaxterBOOK, GaudinBOOK, KorepinBOOK,
  1988_Sklyanin_JPhysA_21_2375}.  In the fermionic language the anisotropy $\Delta$ acts as the
strength of nearest-neighbor interactions.

A particular focus of this work will be on quenches starting from the purely antiferromagnetic regime ($\Delta_i\to\infty$). 
In the limit $\Delta_i\to\infty$ the initial Hamiltonian effectively becomes the Ising Hamiltonian,
$H_i\sim H_{\rm Ising} = \sum_{j=1}^{L} S_j^z S_{j+1}^z$. The ground state of $H_{\rm Ising}$ is exactly
degenerate, with both symmetric and antisymmetric combinations of the N\'eel states having the same
energy. For $\Delta_i$ large but finite this degeneracy is lifted. We denote the two N\'eel states
as
\begin{equation}
  |N_1\rangle = \left|\uparrow\downarrow\right\rangle^{\otimes L/2} \qquad\text{and}\qquad
  |N_2\rangle = \left|\downarrow\uparrow\right\rangle^{\otimes L/2} \epp
\label{eq:Neel_states}
\end{equation}
For $\Delta_i=\infty$, any linear combination of $|N_1\rangle$ and $|N_2\rangle$ is a valid ground
state and, in principle, any such linear combination could be chosen as the initial state with which
to calculate overlaps.  Physically, we find it most reasonable to choose the symmetric combination 
for $L/2$ even and the antisymmetric combination for $L/2$ odd, which is
adiabatically connected to the (unique) ground state for $\Delta_i$ large but finite,
\begin{equation}
  |\Psi_\sigma\rangle = \frac{1}{\sqrt{2}}\left(|N_1\rangle + \sigma |N_2\rangle \right) \qquad\textnormal{with}\quad \sigma = (-1)^{L/2}\epp
  \label{eq:initial}
\end{equation}
For the periodic chain we set $\sigma=\e^{\i p_0}$ with $p_0=0,\pi$. Since $\hat{T}|\Psi_\sigma\rangle =
\e^{\i p_0}|\Psi_\sigma\rangle$, where $\hat{T}$ is the translation operator by one lattice site, $p_0$
is nothing but the momentum of the initial state. For the open chain, $\sigma$
is related to the symmetry of $|\Psi_\sigma\rangle$ under reflection, $\mathscr{R}|\Psi_\sigma\rangle = \sigma
|\Psi_\sigma\rangle$, where the reflection operator $\mathscr{R}$ maps lattice site $j$ to $L+1-j$ for
all $j=1,\ldots,L$.

Recently, some progress has been made in calculating overlaps using the Bethe ansatz \cite{
  2014_Pozsgay_JStatMech_P06011, 2014_Brockmann_JPA_47_145003, 2014_Brockmann_JPA_47_345003,
  2014_Piroli_JPhysA_47_385003}.  There are now exact formulas for the overlap between a N\'eel
state and any eigenstate in terms of the Bethe roots describing the eigenstate
\cite{2014_Brockmann_JPA_47_145003, 2014_Brockmann_JPA_47_345003, 2014_Pozsgay_JStatMech_P06011,
  2012_Kozlowski_JStatMech_P05021, 1998_Tsuchiya_JMathPhys_39_5946}.  By taking the thermodynamic
limit of the results of Ref.~\cite{2014_Brockmann_JPA_47_145003}, it was possible to investigate the
long-time limit in quenches to the antiferromagnetic gapped phase $\Delta_f>1$
\cite{2014_Wouters_PRL_113_117202, 2014_Pozsgay_PRL_113_117203, 2014_Brockmann_JStatMech_P12009} and
also to the isotropic point $\Delta_f=1$ \cite{2014_Brockmann_JStatMech_P12009}, using the quench
action approach \cite{2013_Caux_PRL_110_257203}. However, extracting all the overlaps for large but
finite chains by means of the exact formulas of Refs.~\cite{2014_Brockmann_JPA_47_145003,
  2014_Brockmann_JPA_47_345003} is technically quite challenging.  For this reason, in this work we
will use these forumlas only in the limit $\Delta_f \to 0$, where the structure of Bethe roots
is well understood.

\section{Features of the overlap distribution: numerical overview}\label{sec:Numerical}

In this section we present an overview of the features of overlap distributions for quenches from
various $\Delta_i$ to various $\Delta_f$, obtained using numerical exact diagonalization.  We
present overlaps with only those eigenstates of the post-quench Hamiltonian $H_f$ that lie in the
zero magnetization sector. Since the initial states are always in this sector, eigenstates of $H_f$
in other magnetization sectors have zero overlap due to symmetry.

In cases where eigenstates of $H_f$ are (nearly or exactly) degenerate, numerical diagonalization
might give overlaps with arbitrary linear combinations of the (nearly) degenerate eigenstates.  In
all cases where we discuss visible features of the overlap distribution, we have taken care to
resolve such degeneracies.  There are more degeneracies for periodic chains, so we
show mainly open-chain data.  We have checked in each case that the qualitative features are very
similar for open and periodic chains.

\subsection{Quench from \texorpdfstring{$\Delta_i=\infty$}{} to \texorpdfstring{$\Delta_f=0$}{}}\label{sec:infty_to_zero}

\begin{figure}[tbph]
\centering
\includegraphics[width=0.6\textwidth]{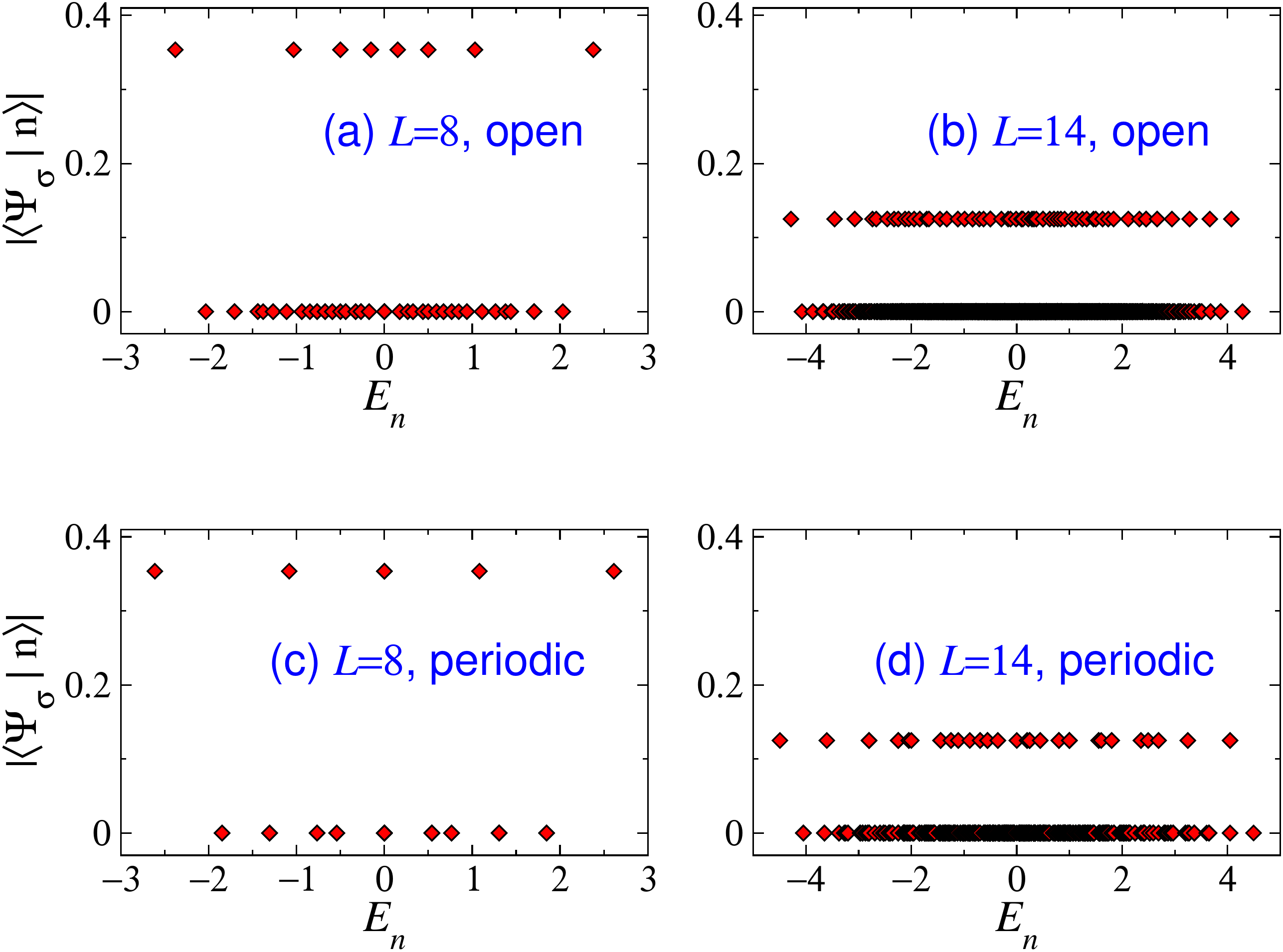}
\caption{Overlap distributions for the quench $\Delta_i=\infty \to \Delta_f=0$.  Data are shown for
  both open and periodic chains, for two different system sizes.  There are $2^{L/2-1}$ nonzero
  overlaps, all exactly equal to $2^{-L/4+1/2}$ which is $2^{-3/2}\approx 0.354$ for $L=8$ and
  $2^{-3}=0.125$ for $L=14$.  In the periodic cases some of the nonzero overlaps are not visually
  distinguishable because of degeneracies, e.g., in panel (c) there are four degenerate nonzero
  overlaps at $E_n=0$.}
\label{fig:infty2zero}
\end{figure}

As explained in Sec.~\ref{sec:model}, for quenches starting from $\Delta_i=\infty$ we choose as the
initial state $\ket{\Psi_0}$ the combination $\ket{\Psi_\sigma}$  of N\'eel states as defined in Eq.~\eqref{eq:initial}.
Overlap distributions for $\Delta_i=\infty$ to $\Delta_f=0$ quenches are shown in
Fig.~\ref{fig:infty2zero}, for both open and periodic chains.  The remarkable feature of these
distributions is that the overlaps are either zero or have exactly equal nonzero value.  
There is no reason to expect that the  ground state at $\Delta=\infty$ should have any  particular preference
between low-energy and high-energy eigenstates of the $\Delta=0$ Hamiltonian. 
Nevertheless, such a physical argument does not suffice to predict that the nonzero overlaps should
all be exactly equal.  This is a particular feature of the N\'eel-to-XX quench in the XXZ chain.

Among the $\binom{L}{L/2}$ final eigenstates in the zero magnetization sector, only
$2^{L/2-1}$ have nonzero $c_n$'s. The number of nonzero overlaps fixes the nonzero values to be
$|c_n|=2^{-L/4+1/2}$, by virtue of normalization and the fact that they are all
equal. The counting can sometimes be tricky due to degeneracies. For example, there appears to be only five
instead of eight nonzero overlaps in the $L=8$ periodic case in Fig.~\ref{fig:infty2zero}(c).
This is because the symbol at the center represents four exactly degenerate eigenstates, which cannot
be distinguished.  There are also few eigenstates with the same eigenenergy but zero overlap. Of
course, any linear combination of the degenerate eigenstates is also a valid eigenstate. So, there is some
ambiguity in the overlaps plotted.  By choosing the eigenstates to be the appropriate `Slater
determinants' in fermionic language (see Sec.~\ref{sec:B}), the overlaps turn out to be either
zero or $|c_n|=2^{-L/4+1/2}$.

The overlap distribution for quenches from $\Delta_i=\infty$ to $\Delta_f=0$ will be analyzed in
detail in Sec.~\ref{sec:B}, in particular we will derive the selection rules, which determine whether
or not an eigenstate has nonzero overlap.  We will also exploit this analysis to obtain closed
expressions for time-dependent quantities.

\subsection{Quenches from  \texorpdfstring{$\Delta_i=\infty$}{}  to various nonzero \texorpdfstring{$\Delta_f$}{}} \label{sec:numeric_infty2various}

We  now consider quenches starting from the $\ket{\Psi_\sigma}$ state ($\Delta_i=\infty$) to various
nonzero values of $\Delta_f$.  
%
\begin{figure}[tb]
\centering
\includegraphics[width=0.8\textwidth]{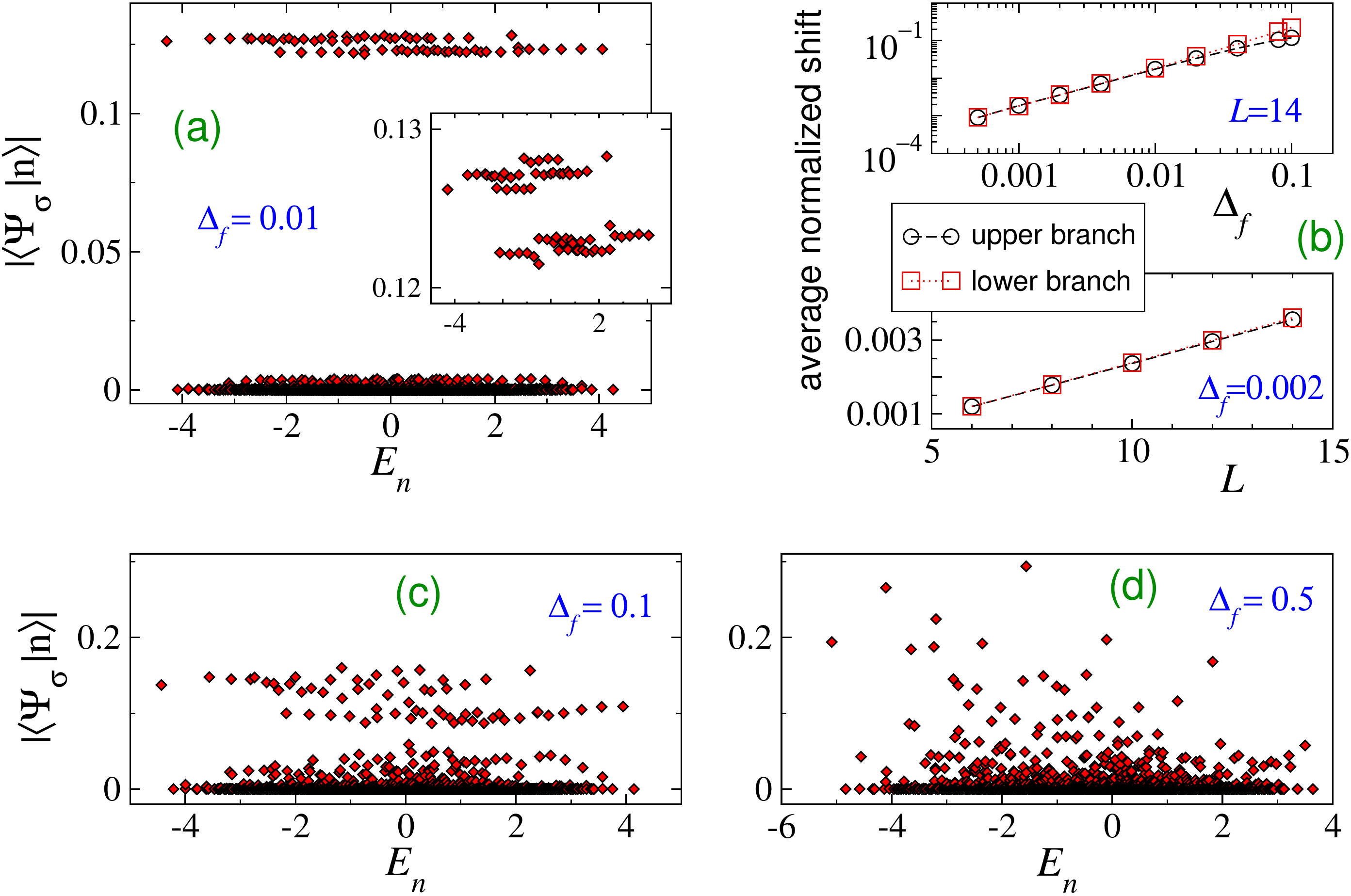}
\caption{(a,c,d) Overlap distributions for $\Delta_i=\infty$ (initial state
  $\ket{\Psi_0}=\ket{\Psi_\sigma}$) in an open chain with $L=14$ sites. The splitting that
  can be observed for $\Delta_f\ll1$ disappears as $\Delta_f$ grows.  (b) Magnitude of the average
  normalized shifts for the two branches as function of $\Delta_f$ (double logarithmic scale) and as
  function of system size $L$ (linear scale).  The normalized shift for small $\Delta_f$ is observed
  to be linear in $\Delta_f$ and $L$.}
\label{fig:smalldelta}
\end{figure}
%
In Fig.~\ref{fig:smalldelta}(a,c,d), we show overlap distributions for quenches to $\Delta_f$ values
in the gapless region, $\Delta_f<1$.  When $\Delta_f$ is very small, the distribution is only
slightly distorted from the $\Delta_f=0$ case, as expected from perturbative arguments.
However, the pattern of distortion is rather remarkable.  The $2^{L/2-1}$ nonzero overlaps now split into two
well-defined branches; one branch is shifted upwards and one branch is shifted downwards, see
Fig.~\ref{fig:smalldelta}(a).  The upward- (resp.~downward-) shifted branch is biased toward lower
(resp.~higher) energies.  In particular, among the eigenstates satisfying the selection rule,
the lowest few all have positive shift, the highest few always have negative shift, and the
eigenstates with intermediate energies can have either positive or negative shift.  In
addition, many of the overlaps that were zero for the $\Delta_f=0$ case, now have nonzero (albeit
tiny) $|c_n|$ values.

We will refer to the difference $|c_n|-2^{-L/4+1/2}$  of the absolute 
value of the N\'eel overlap with an eigenstate of the
$\Delta_f \ll 1$ Hamiltonian and the N\'eel overlap at $\Delta_f=0$, i.e., $2^{-L/4+1/2}$, 
as the ``shift'' of the overlap.  The splitting structure
is well-defined as long as the shift is much smaller than $2^{-L/4+1/2}$; otherwise the lower branch is
difficult to distinguish from the states with near-zero overlap. It is useful to normalize the shift
with respect to the $\Delta_f=0$ value, i.e., for each eigenstate connected to one having nonzero
overlap at  $\Delta_f=0$, we define 
\begin{equation}  \label{eq:normalized_shift_defn}
  \text{normalized shift} = \frac{\text{shift}}{2^{-L/4+1/2}} = \frac{|c_n|-2^{-L/4+1/2}}{2^{-L/4+1/2}} \epp
\end{equation}
In Fig.~\ref{fig:smalldelta}(b), the magnitudes of the normalized shifts are averaged over the eigenstates in each
branch.  We see this to be linear with $\Delta_f$ at fixed $L$, and linear with $L$ at fixed
$\Delta_f$, i.e., proportional to $L\Delta_f$.  This indicates that the shift can be calculated
perturbatively in $\Delta_f$ for any finite chain. However as the chain size gets larger the perturbative region
of $\Delta_f$ shrinks.  In Sec.~\ref{sec:A} we will analyze the splitting
and shift using first-order perturbation theory.  The branches have internal structures, e.g., the
overlap values roughly increase with energy within each branch.  

For a fixed $L$, as $\Delta_f$ increases further, the splitting structures gets diffuse, and the
separation between the $2^{L/2}$ larger values and the rest smaller values also gets obscured,
see Fig.~\ref{fig:smalldelta}(c).  Eventually, there is no visible remnant of the characteristic
structure at $\Delta_f=0$, as seen for $\Delta_f=0.5$ in Fig.~\ref{fig:smalldelta}(d).  For
$\Delta_f=0.5$ the overlap distribution has no striking pattern, and no obvious remnant of the flat
distribution at $\Delta_f=0$.  A slight bias toward smaller energies is visible.  This may be
expected because, for any finite $\Delta_f>0$, the $\Delta_i=\infty$ ground state
$\ket{\Psi_\sigma}$ should have more weight in the lower part of the spectrum than in the upper
part.  

\begin{figure}[tbph]
\centering
\includegraphics[width=0.8\textwidth]{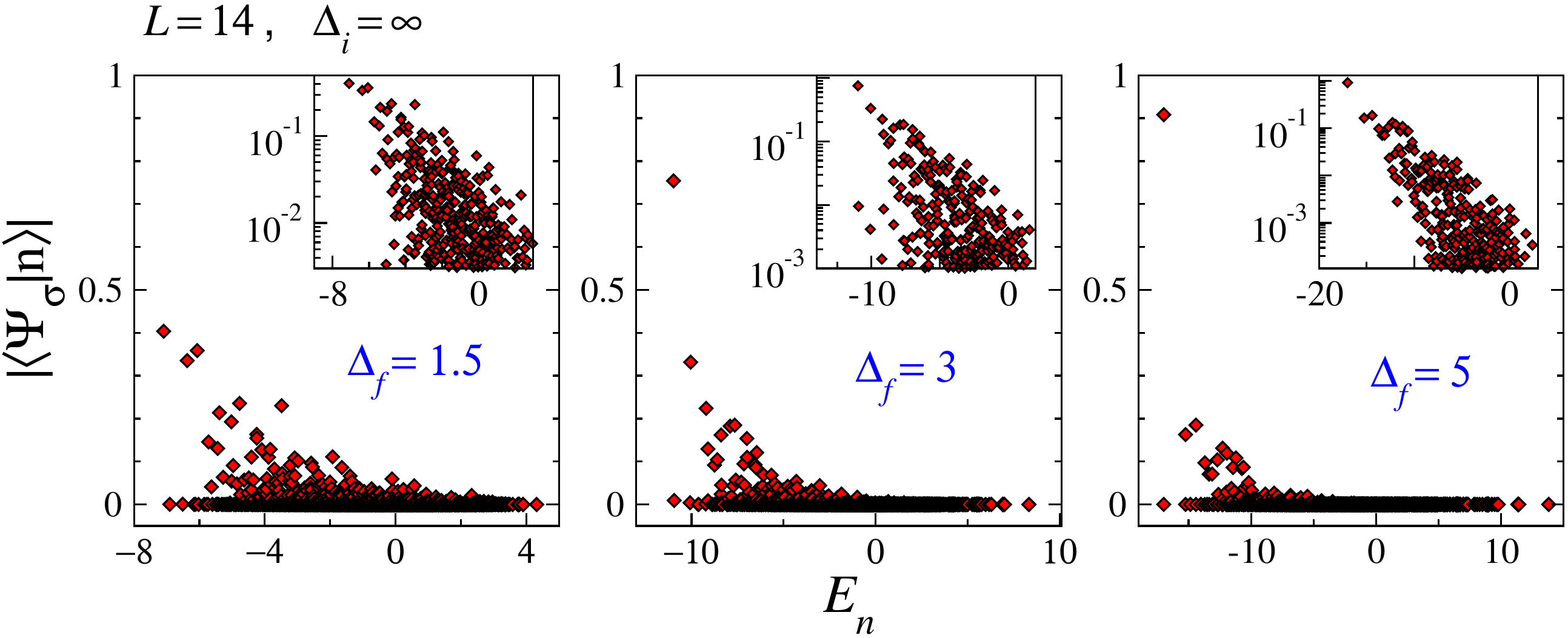}
\caption{ Overlap distributions for $\Delta_i=\infty$ (initial state
  $\ket{\Psi_0}=\ket{\Psi_\sigma}$) and $\Delta_f > 1$ in an open chain with $L=14$ sites. The
  overlap distributions decay with increasing energy. The insets show the same data in a logarithmic
  scale.}
\label{fig:largedelta}
\vspace{5ex}
\includegraphics[width=0.99\textwidth]{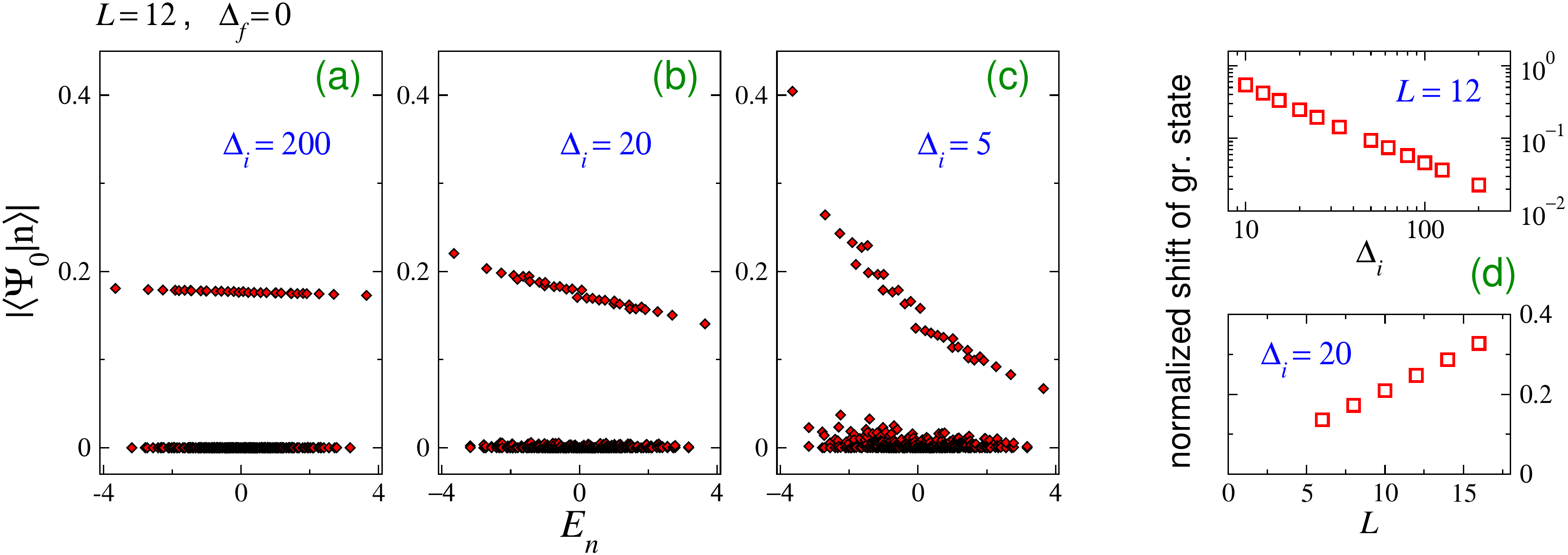}
\caption{ (a,b,c) Overlap distributions for $\Delta_i\gg 1$ and $\Delta_f=0$ in an open chain with
  $L=12$ sites.  (d) The normalized shift of the lowest-energy eigenstate, as a function $\Delta_i$
  for fixed chain length $L$ (double logarithmic scale), and as a function of chain
  length $L$ for fixed $\Delta_i$ (linear scale).  
}
\label{fig:fromlarge}
\vspace{5ex}
\includegraphics[width=0.9\textwidth]{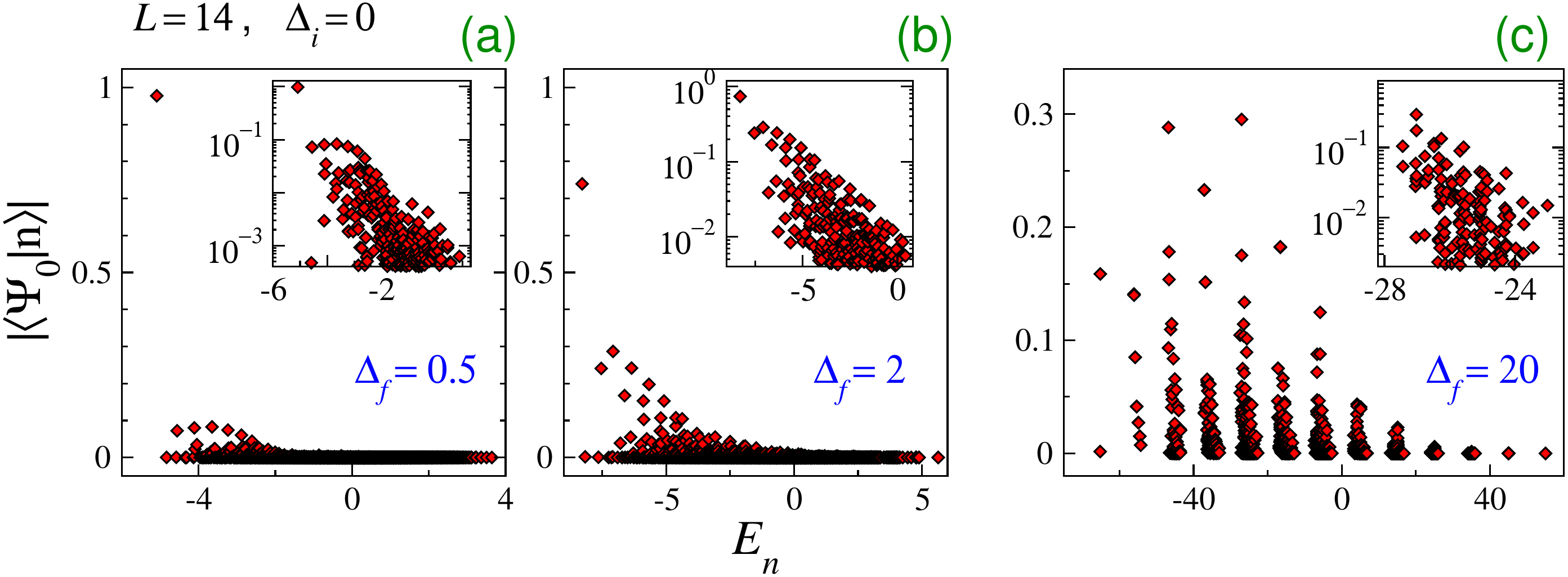}
\caption{Overlap distributions for $\Delta_i=0$ and various $\Delta_f$ in an open chain with $L=14$
  sites. In the last panel we see how for large values of $\Delta_f$ the energy spectrum is
  organized in `bands'.}
\label{fig:zero2various}
\end{figure}

This trend gets more prominent at larger $\Delta_f$.  In Fig.~\ref{fig:largedelta} we show some
examples for $\Delta_f>1$.  The bias toward lower-energy states gets progressively stronger.  This
indicates that for quenches from $\Delta_i=\infty$ to $\Delta_f>1$, the dynamical evolution will be
determined mostly by low-energy states.  For $\Delta_f \gtrsim 3$ the ground state dominates the
overlap distribution very strongly.

\subsection{Quenches from \texorpdfstring{$\Delta_i \gg 1$}{} to \texorpdfstring{$\Delta_f=0$}{}}

We now consider quenches to $\Delta_f=0$, starting from $\Delta_i \gg 1$ large but finite.  The
initial state $|\Psi_0\rangle$ is now no longer exactly the combination $\ket{\Psi_\sigma}$ of N\'eel states.  From
perturbative intuition, we expect that as $\Delta_i$ decreases toward smaller values the overlap
distribution will deviate gradually from the characteristic shape of Fig.\ \ref{fig:infty2zero}.
This is seen to be true in Fig.~\ref{fig:fromlarge}, where we have plotted overlap distributions for
an open chain with $L=12$, for three different $\Delta_i \gg 1$. The $2^{L/2-1}$ nonzero values exhibit a
roughly linear dependence on $E_n$ as $\Delta_i$ is decreased from $\infty$.  One expects the ground
state at any finite $\Delta_i$ to be closer to the low-energy eigenstates of the $\Delta_f=0$ than
to the higher-energy eigenstates. Hence, it is not surprising that the slope of the line is negative.
The set of overlaps fall on a very nearly linear curve for large $\Delta_i$. For moderate
$\Delta_i$, see Fig.~\ref{fig:fromlarge}(c), the overlaps follow a more curved line and show more
scatter.  In addition the near-zero overlap values start to be noticeable.

The shift of the ground state overlap from $2^{-L/4+1/2}$ can be seen numerically to scale as
$\propto1/\Delta_i$ for a given $L$ and large $\Delta_i$, see Fig.~\ref{fig:fromlarge}(d).  The overlap
of the highest-energy eigenstate obeying the selection rules scales in the same way.  Also, the
normalized shift for the ground state scales linearly with chain length $L$ for a given $\Delta_i$.

In Sec.~\ref{sec:C}, treating the XX part of the Hamiltonian as a perturbation to the Ising
Hamiltonian, we will show that the shift of the overlaps is exactly a linear function of energy at
first order in $1/\Delta_i$.  In fact, the normalized shift for the eigenstates satisfying the
selection rule will be shown to be $-E_n/\Delta_i$ at first order (exact for periodic chains,
approximate up to order $1/L$ for open chains), consistent with the numerical data.

\subsection{Quenches from \texorpdfstring{$\Delta_i=0$}{} to various \texorpdfstring{$\Delta_f$}{}}

We now consider quenches starting from $\Delta_i=0$. Thus, the initial state is
the ground state of a free fermionic Hamiltonian. Examples for three $\Delta_f$ values are shown in
Fig.~\ref{fig:zero2various} for an $L=14$ open chain.  For any finite value of $\Delta_f$, we
expect a bias toward low-energy states. This is seen in all three cases.  The bias is naturally
stronger for smaller $\Delta_f$. We see that up to $\Delta_f=2$ the ground state overlap dominates
the overlap distribution.  Other than the ground state, most of the weight for $\Delta_f\approx 1$ 
goes to a few final eigenstates at relatively low energies.  We also see from the logarithmic plots in
the insets of Fig.~\ref{fig:zero2various}(a,b) that there is a characteristic decaying shape.  The
upper envelope of this scatter-plot decays roughly exponentially, although a sharp qualitative statement
about the decay rate is difficult to make, due to finite-size effects.

The $\Delta_f\gg1$ case has some bunching structure reflecting the fact that the energy spectrum near the
Ising limit is grouped into `bands'.  Each band is characterized by the number of domain walls in the
configurations that dominate the eingenstates of that band.  Accordingly, the overlap distribution
is broken into groups.  Interestingly, within each band the overlap distribution is dominated by the
lower-energy part of that band. Moreover, each band of the distribution has roughly similar shape as the
characteristic shape of the full distribution for the $\Delta_f\approx 1$ case (see Figs.~\ref{fig:zero2various}(a,b)).
The inset to Fig.~\ref{fig:zero2various}(c) exemplifies this by zooming into one of the bands.

\section{Analytical results for the quench from \texorpdfstring{$\Delta_i=\infty$}{} to \texorpdfstring{$\Delta_f=0$}{}}\label{sec:B}

In this section, we will present a series of analytical results on the quench from $\Delta_i=\infty$
to $\Delta_f=0$, i.e., the quench from the combination $\ket{\Psi_\sigma}$ of N\'eel states to the XX
chain.  This is the case addressed numerically in Sec.~\ref{sec:infty_to_zero} and
Fig.~\ref{fig:infty2zero}.  In Sec.~\ref{sec:freefermions}, we set notation by providing a brief 
reminder of how the XX Hamiltonian is diagonlized by mapping to Jordan-Wigner fermions.

In Sec.~\ref{sec:selection}, we use the fermionic language to derive selection rules for an XX
eigenstate to have nonzero overlap with $\ket{\Psi_\sigma}$.  The rules are expressed in terms of
the momenta (periodic chain) or pseudo-momenta (open chain) of the fermions.  Physically, one rule
arises from the structure of the N\'eel states $\ket{N_{1,2}}$, and a second rule arises from the
well-defined symmetry of the combination $\ket{\Psi_\sigma}$.   We also express the overlap
magnitudes as a sum of determinants (one for each $\ket{N_{1,2}}$).  This is evaluated as a function
of the chain length $L$, both by exploiting determinant properties, and from normalization, using
the fact that all the nonzero overalps are equal for this quench.  

In Sec.~\ref{sec:ba}, we use the overlap formulas of Refs.~\cite{2014_Brockmann_JPA_47_145003,
  2014_Brockmann_JPA_47_345003} (valid for all $\Delta_f$) and take the limit $\Delta_f\to 0$.  In
this way we formulate selection rules in the language of Bethe roots.  We show how these rules are
equivalent to the rules in terms of fermionic (pseudo-)momenta.

Finally, we provide two applications of our detailed understanding of the overlap distribution in
this particular case.  In Sec.~\ref{sec:LoschEch}, we derive closed expressions for the Loschmidt
echo after the quench.  In Sec.~\ref{sec:correlators}, we present closed expressions for the full
time dependence of equal-time two-point correlators.  We consider both longidudinal
($\left<S_{j}^zS_{j+n}^z\right>$) and transverse ($\left<S_{j}^+S_{j+n}^-\right>$) correlators. The
expressions for the transverse correlators appear here for the first time, to the best of our
knowledge.

\subsection{XX eigenstates in free-fermion language}\label{sec:freefermions}

To diagonalize $H_{\text{xx}}$, i.e., the $\Delta=0$ part of the Hamiltonian \eqref{eq:ham}, one uses the
Jordan-Wigner (JW) transformation 
\begin{equation}\label{eq:JW}
  S_j^+ = S_j^x + \i S_j^y = c_j^\dag\exp\left(\i\pi\sum_{i=1}^{j-1}c_i^\dag c_i^{\phantom\dag}\right)\epc \qquad S_j^z=c_j^\dag c_j^{\phantom\dag} -\frac{1}{2}\epp
\end{equation}
Here $c_j^\dag$ and $c_j^{\phantom\dag}$ are usual fermionic operators, satisfying 
anticommutation relations $\{c_i^{\phantom\dag},c_j^\dag\}=\delta_{ij}$,
$\{c_i^\dag,c_j^\dag\}=\{c_i^{\phantom\dag},c_j^{\phantom\dag}\}=0$.

For periodic boundary conditions, the JW transformation turns $H_{\text{xx}}$ to the free fermionic
Hamiltonian $H^{\rm PBC} = \frac{1}{2}\sum_{j=1}^{L}\left(c^\dag_{j+1}c_j^{\phantom\dag} + c_j^\dag
  c_{j+1}^{\phantom\dag}\right)$, with the boundary condition $c_{L+1}^{\phantom\dag}=\e^{\i\pi
  \sum_{j=1}^{L}c_j^\dag c_j^{\phantom\dag}} c_1^{\phantom\dag} = (-1)^{\hat N} c_1^{\phantom\dag}$,
where $N = \langle \hat N \rangle$ is the number of fermions in the state.  Using a Fourier
transformed set of operators,
\begin{equation}\label{eq:Fourier}
  \tilde{c}_k^\dag=\frac{1}{\sqrt{L}}\sum_{j=1}^{L}\e^{\i kj}c_j^\dag \epc
\end{equation}
$H^{\rm PBC}$ can be written in diagonal form
\begin{equation}
  H^{\rm PBC} = \sum_{k}\epsilon_k\tilde{c}^\dag_k\tilde{c}_k^{\phantom\dag}\epc \quad \epsilon_k = \cos(k) \epp
  \label{eq:ham_pbc}
\end{equation}
The boundary conditions enforce  quantization of the momenta $k$: 
\begin{equation}\label{eq:set_allowed_momenta}
  k \in K_a = \left\{\frac{2\pi(m-a)}{L}\right\}_{m=1}^{L} \epc
\end{equation}
where $a=1/2$ for $N$ even, and $a=0$ for $N$ odd.

For the open chain, the JW transformation leads to $H^{\rm
  OBC}=\frac{1}{2}\sum_{j=1}^{L-1}(c^\dag_{j+1}c_j^{\phantom\dag} + c_j^\dag
c_{j+1}^{\phantom\dag})$.  This Hamiltonian is diagonalized be introducing the canonical operators 
\begin{equation}
  \bar{c}_q^\dag = \sqrt{\frac{2}{L+1}}\sum_{j=1}^L\sin(qj)c^\dag_j \epp
  \label{eq:Fouriersine}
\end{equation}
leading to 
\begin{equation}
  H^{\rm OBC} = \sum_{q}\epsilon_q\bar{c}^\dag_q\bar{c}_q^{\phantom\dag} \epc \quad  \epsilon_q = \cos(q)\epc
  \label{eq:free}
\end{equation}
provided that the numbers $q$ are quantized as
\begin{equation}
  q \in Q = \left\{\frac{\pi m}{L+1}\right\}_{m=1}^{L}\epp
  \label{eq:quantization}
\end{equation}
Since there is no translational invariance in the open chain, we call $q$ ``pseudo-momenta''.

\subsection{Determinant formulas and selection rules}\label{sec:selection}

In terms of the Jordan-Wigner fermions introduced above, our initial state \eqref{eq:initial} can be
written as
\begin{equation}\label{eq:initial_by_cs}
  |\Psi_\sigma\rangle = \frac{1}{\sqrt{2}}\left(\prod_{j=1}^{L/2} c_{2j-1}^\dag + \sigma\prod_{j=1}^{L/2} c_{2j}^\dag\right)|0\rangle \epc
\end{equation}
where $|0\rangle$ is the vacuum state (all spins down).  The eigenstates of the XX Hamiltonian in
the zero-magnetization ($N=L/2$ fermion) sector can be written for the periodic (resp.\ open) chain
as
\begin{equation}\label{eq:free_fermion_state}
  |\{k_\a\}_{\a=1}^{L/2}\rangle = \tilde{c}_{k_1}^\dag \tilde{c}_{k_2}^\dag \dots \tilde{c}_{k_{L/2}}^\dag |0\rangle \quad\left(\text{resp.}\ |\{q_\a\}_{\a=1}^{L/2}\rangle = \bar{c}_{q_1}^\dag \bar{c}_{q_2}^\dag \dots \bar{c}_{q_{L/2}}^\dag |0\rangle\right)
\end{equation}
with $k_\a\in K_a$ (resp.~$q_\a \in Q$) for all $\a = 1,\ldots,L/2$, and $k_\a \neq k_\beta$ (resp.~$q_\a \neq q_\beta$) for $\a\neq\beta$.

The overlap between an eigenstate of the form \eqref{eq:free_fermion_state} and the initial state
$\ket{\Psi_\sigma}$ from Eq.~\eqref{eq:initial_by_cs} can now be computed using Wick's theorem. For
example, with periodic boundary conditions, we obtain
\begin{align}\label{eq:wick1}
 \langle \Psi_\sigma|\{k_\a\}_{\a=1}^{L/2}\rangle &= \frac{1}{\sqrt{2}}\left[\langle 0 |c_1^{\phantom{\dag}} c_3^{\phantom{\dag}}\ldots c_{L-1}^{\phantom{\dag}} \tilde{c}_{k_1}^\dag  \tilde{c}_{k_2}^\dag \dots \tilde{c}_{k_{L/2}}^\dag|0\rangle + \sigma \langle 0|c_2^{\phantom{\dag}} c_4^{\phantom{\dag}} \ldots c_{L}^{\phantom{\dag}} \tilde{c}_{k_1}^\dag  \tilde{c}_{k_2}^\dag \dots \tilde{c}_{k_{L/2}}^\dag|0\rangle\right] \\ \label{eq:wick2}
 &= \frac{1}{\sqrt{2}}\left[\det_{1\leq j,\a\leq L/2}\left(\langle0|c_{2j-1}^{\phantom{\dag}}\tilde{c}_{k_\a}^\dag|0\rangle\right) + \sigma\det_{1\leq j,\a\leq L/2}\left(\langle 0|c_{2j}^{\phantom{\dag}}\tilde{c}_{k_\a}^\dag|0\rangle\right)\right]\epp
\end{align}
Equation \eqref{eq:wick2} follows from Eq.~\eqref{eq:wick1} by separate application of Wick's theorem on the two correlators. 
Using definition \eqref{eq:Fourier} this can be simplified to
\begin{align}
 \langle\Psi_\sigma|\{k_\a\}_{\a=1}^{L/2}\rangle &= \frac{1}{\sqrt{2}} \left(\frac{1}{L}\right)^{L/4}\left[\det_{1\leq j,\a\leq L/2}\Big(\e^{\i k_\a(2j-1)}\Big)  + \sigma\det_{1\leq j,\a\leq L/2}\Big(\e^{\i k_\a 2j}\Big)\right] \epp
  \label{eq:overlapPBC}
\intertext{For the open chain we similarly obtain}
\label{eq:overlapOBC}
  \langle \Psi_\sigma|\{q_\a\}_{\a=1}^{L/2}\rangle &= \frac{1}{\sqrt{2}}\left(\frac{2}{L+1}\right)^{L/4}\left[\det_{1\leq j,\a\leq L/2}\Big(\sin\big(q_\a(2j-1)\big)\Big) + \sigma\det_{1\leq j,\a\leq L/2}\Big(\sin\big(2q_\a j\big)\Big)\right] \epp
\end{align}
The overlaps \eqref{eq:overlapPBC} and \eqref{eq:overlapOBC} can be zero if both determinants vanish, or if their sum exactly cancels. As discussed below, it turns out that both cases can occur. 

We first analyze the formula for the periodic chain. We set $\sigma=\e^{\i p_0}$ and simplify the
square bracket in Eq.~\eqref{eq:overlapPBC},
\begin{equation}
  \det_{1\leq j,\a\leq L/2}\Big(\e^{\i k_\a(2j-1)}\Big) + \sigma\det_{1\leq j,\a\leq L/2}\Big(\e^{2\i k_\a j}\Big) = \det_{1\leq j,\a\leq L/2}\Big(\e^{2\i k_\a j}\Big)\left(\e^{-\i\sum_{\a=1}^{L/2} k_\a} +\e^{\i p_0}\right) \epp
  \label{eq:overlapPBC2}
\end{equation}
Now, if there is a pair of the form $\{k,k+\pi\}\subset \{k_\a\}_{\a=1}^{L/2}$, the determinant in
Eq.~\eqref{eq:overlapPBC2} is zero as two columns of the matrix are equal,
$\e^{2i(k+\pi)j}=\e^{2ikj}$ for all $j$. The set $K_a$ of all possible momenta, defined in Eq.~\eqref{eq:set_allowed_momenta}, decomposes into $L/2$ many of such pairs: 
\begin{equation}
  K_a = \bigcup_{m=1}^{L/2} \left\{\frac{2\pi(m-a)}{L},\frac{2\pi(m-a)}{L}+\pi\right\} \epc
\end{equation}
with  $a=1/2$  for $L/2$ even and $a=0$ for  $L/2$  odd. In order to get a non-vanishing determinant, 
each pair in this decomposition has to be occupied by
exactly one momentum. This shows that there are only $2^{L/2}$ states for which the determinant can
be nonzero. If we flip one of the momenta by $\pm\pi$, i.e., choosing the other member of the pair
(`pair flip'), the determinant does not change, again due to $\e^{2\i (k\pm\pi)j}=\e^{2\i
  kj}$. Therefore, the determinant has always the same non-vanishing value for all of these
$2^{L/2}$ states. Note that the overlaps \eqref{eq:overlapPBC} might differ by a factor $-1$ due to
the different order of momenta in pair-flipped states. 

The other momentum-dependent factor in Eq.~\eqref{eq:overlapPBC2} is $\e^{-\i\sum_{\a=1}^{L/2}
  k_\a}+\e^{\i p_0}$, which is zero if the total momentum $\sum_{\a=1}^{L/2} k_\a$ of the eigenstate
$|\{k_\a\}_{\a=1}^{L/2}\rangle$ and the momentum $p_0$ of the initial state $|\Psi_\sigma\rangle$
differ by $\pi$ (modulo $2\pi$). This means that only an even number of pair flips are
allowed.  Hence, the number of states that have nonzero overlap gets reduced by a factor of $2$.

Thus, there are in total $2^{L/2-1}$ eigenstates $|\{k_\a\}_{\a=1}^{L/2}\rangle$, all with the same
overlap (up to a possible global minus sign as dicussed above). Since the initial state
$\ket{\Psi_\sigma}$ is normalized, this common value is
\begin{equation}
  \left|\langle\Psi_\sigma|\{k_\a\}_{\a=1}^{L/2}\rangle\right| = 2^{-L/4+1/2}\epp
\end{equation}

The derivation above did not require any explicit determinant calculation.  We also checked that the
result can be recovered by directly computing the determinant in Eq.~\eqref{eq:overlapPBC2}. The
Vandermonde determinant formula $\det_{1\leq j,l\leq n}\left(x_j^{l-1}\right)=\prod_{1\leq j<l\leq
  n}\left(x_j-x_l\right)$ allows to express it as a double product that can be further
simplified. Crucial in the evaluation is the formula $\prod_{p=1}^{n-1}(1-\e^{2ip\pi/n})=n$, which
can be established by noticing that $\prod_{p=1}^{n-1}(z-\e^{2ip\pi/n})=\frac{z^n-1}{z-1}$ and
$\lim_{z\to 1} \frac{z^n-1}{z-1}=n$.

The analysis of the overlap formula \eqref{eq:overlapOBC} for the open chain is similar. If there is
a pair of the form $\{q,\pi-q\}\subset \{q_\a\}_{\a=1}^{L/2}$, both determinants are zero as two
columns are proportional to each other: $\sin(2(\pi-q)j)=-\sin(2qj)$ or
$\sin((\pi-q)(2j-1))=\sin(q(2j-1))$, respectively. The set $Q$ of all possible pseudo-momenta, see Eq.~\eqref{eq:quantization},
decomposes into $L/2$ many of such pairs:
\begin{equation}
  Q = \bigcup_{m=1}^{L/2} \left\{\frac{\pi m}{L+1},\pi -\frac{\pi m}{L+1}\right\} \epp
\end{equation}
Again, in order to get non-vanishing determinants, each pair in this decomposition has to be
occupied by exactly one pseudo-momentum. If we flip one of them by $\pm\pi$, i.e., doing a pair flip,
the second determinant does not change, while the first one picks up a minus sign. This shows that
there are again $2^{L/2}$ states for which the determinants are nonzero and that they all have the
same absolute value. The two determinants are equal for the state where all pseudo-momenta are
smaller than $\pi/2$ and also for all states that can be obtained from this state by an even number
of pair flips. In contrast, for states obtained by an odd number of pair flips they differ by a
factor $-1$. If $n$ denotes the number of pair flips, i.e., the number of pseudo-momenta larger than
$\pi/2$, the overlap is proportional to $(-1)^{n}+\sigma$, which is zero for $\sigma=1$ and $n$ odd
or $\sigma=-1$ and $n$ even, and it is $\pm 2$ otherwise. This supplemental restriction also follows from
the reflection symmetry of the initial state. Since the reflection operator $\mathscr{R}$ commutes
with the Hamiltonian, the energy eigenstates are also eigenstates of $\mathscr{R}$,
i.e.~$\mathscr{R}|\{q_\a\}_{\a=1}^{L/2}\rangle = r|\{q_\a\}_{\a=1}^{L/2}\rangle$. Hence,
\begin{equation}
  \sigma \langle\Psi_\sigma|\{q_\a\}_{\a=1}^{L/2}\rangle = \langle\Psi_\sigma|\mathscr{R}|\{q_\a\}_{\a=1}^{L/2}\rangle = r\langle\Psi_\sigma|\{q_\a\}_{\a=1}^{L/2}\rangle
\end{equation}
and the overlap with $|\Psi_\sigma\rangle$ can only be nonzero if $r=\sigma$. (In the fermionic language
the action of $\mathscr{R}$ on a many particle state can be deduced from the formula
$\mathscr{R}c_j^\dag \mathscr{R}^{-1} = (-1)^{\hat{N}}c_{L+1-j}^\dag$, where $\hat{N}$ is the total
fermion number. Applying this to the N\'eel state we recover the same rule.) Therefore, as in the
periodic case, the number of states that have nonzero overlap gets reduced by a factor of $2$ due to
the reflection symmetry. Again, there are $2^{L/2-1}$ eigenstates with overlap
\begin{equation}
  \left|\langle\Psi_\sigma|\{q_\a\}_{\a=1}^{L/2}\rangle\right| =  2^{-L/4+1/2}\epp
\end{equation}
Note that this result is in agreement with the ground-state overlap of
Ref.~\cite{2011_Stephan_PRB_84_195128}, obtained by an explicit determinant calculation. All other
eigenstates (those with a pair $\{q,\pi-q\}$ or those with the wrong number of momenta larger than
$\pi/2$) have zero overlap with $|\Psi_\sigma\rangle$.

Let us summarize our findings. We have a simple algorithm that allows us to determine which
post-quench eigenstates \eqref{eq:free_fermion_state} have nonzero overlap with the initial state
$\ket{\Psi_\sigma}$ and, therefore, contribute to the time evolution and to the non-equilibrium
dynamics after the quench. For that to happen two conditions need to be fulfilled:
\begin{enumerate}[label=(\roman*)]
 \item \label{rule1}The first condition (`forbidden pairs') selects the states for which the determinants are individually nonzero, i.e.~states with momenta or pseudo-momenta satisfying
\begin{align}
	|k_\alpha - k_\beta| &\neq \pi \quad\textnormal{for all $\alpha,\beta\in\{1,\ldots,L/2\}$ (periodic chain)} \label{eq:selectionrule1}\epc\\
	q_\alpha + q_\beta &\neq \pi \quad\textnormal{for all $\alpha,\beta\in\{1,\ldots,L/2\}$ (open chain)} \epp
\end{align}
\item \label{rule2}The second condition (`number of pair flips') selects the states for which the
  two terms in Eqs.~\eqref{eq:overlapPBC} and \eqref{eq:overlapOBC} do not cancel.  This corresponds
  to selecting eigenstates with the same symmetry (translation or reflection) as the initial state.
  In the periodic case this restriction can be written as $\sum_{\a=1}^{L/2}k_\a = p_0$, which
  relates the total momentum of the eigenstate to the momentum $p_0$ of the initial state.  For the
  open chain the restriction is imposed by the reflection symmetry of the initial state,
  $\mathscr{R}|\Psi_\sigma\rangle = \sigma|\Psi_\sigma\rangle$, via $\sigma = (-1)^{n}$, where $n$
  is the number of pseudo-momenta larger than $\pi/2$.
\end{enumerate}
Note that, if we use $\ket{N_1}$ or $\ket{N_2}$ as initial state, only the first rule \ref{rule1} above
applies and there are $2^{L/2}$ nonzero overlaps.  The second rule \ref{rule2}
arises because $\ket{\Psi_\sigma}$ respects the translation/reflection symmetry of the Hamiltonian.

\subsection{Overlaps derived from a Bethe Ansatz approach}\label{sec:ba}

In this subsection, we treat the overlap distribution for the $\Delta_{i}=\infty$ to $\Delta_{f}=0$ quench
using the Bethe Ansatz formalism.  Recent work has provided expressions for the overlap between
N\'eel states and Bethe eigenstates at arbitrary $\Delta_{f}$
 \cite{2014_Brockmann_JPA_47_145003,
  2014_Brockmann_JPA_47_345003, 2014_Pozsgay_JStatMech_P06011, 2012_Kozlowski_JStatMech_P05021,
  1998_Tsuchiya_JMathPhys_39_5946}.  We specialize the formulas of
Ref.~\cite{2014_Brockmann_JPA_47_345003} to the XX limit ($\Delta_{f}=\cos(\gamma)\to 0$, i.e.~$\gamma\to\pi/2$).  
We will show that the condition of having nonzero overlap with an XX Bethe state
can be visualized by adjacent corners of certain rectangles in the pattern of possible XX Bethe
roots. We further relate this condition to the selection rule \eqref{eq:selectionrule1} of single
particle momenta.  The description here is mostly self-contained, but the interested reader may wish
to consult Refs.\ \cite{2014_Brockmann_JPA_47_145003, 2014_Brockmann_JPA_47_345003} for background
and details on related matters.

For simplicity we focus on the case $L/2$ even.  We start our analysis by presenting the solutions
of XX Bethe equations with twist and constant inhomogeneity (see, e.g., \cite{BaxterBOOK}). The
twist deformation is related to twisted boundary conditions of the spin chain, $S_{L+1}^\a =
\e^{\i\phi} S_1^\a$, $\a=x,y,z$. It is necessary to introduce a nonzero twist to resolve issues with
singular terms in the norm formula for Bethe states in the limit $\gamma\to\pi/2$. We then use
`twist-deformed' solutions to calculate overlaps by means of the aforementioned formulas.

The Bethe equations for anisotropy $\Delta=\cos(\gamma)$ with twist $\phi$ and constant inhomogeneity $\epsilon$ read
\begin{equation}
  \left[\frac{\sinh(\lambda_j - \epsilon +\frac{\i\gamma}{2})}{\sinh(\lambda_j- \epsilon -\frac{\i\gamma}{2})}\right]^L = -\e^{\i\phi} \prod_{j'=1}^{M}\frac{\sinh(\lambda_j -\lambda_{j'}+\i\gamma)}{\sinh(\lambda_j-\lambda_{j'}-\i\gamma)}\epc \qquad j = 1,\ldots, M \epp
\label{eq:BEtwist}
\end{equation}
We consider $\phi,\epsilon \in\mathbb{R}$ small. For $\gamma = \pi/2$ and $M = L/2$ even, the right hand side simplifies to $-\e^{\i\phi}$, which eventually leads to 
\begin{equation}
  \tanh(\lambda_j-\epsilon) = \tan\big(\tfrac{\pi}{L}(n_j-\tfrac{1}{2}) - \tfrac{\phi}{2L}\big)\epc \qquad j=1,\ldots,L/2 \epc
\label{eq:BE_XX_twist}
\end{equation}
where $n_j$ are integers.  For small $\phi$ and $\epsilon$, we have $-L/4+1 \leq n_j \leq L/4$ for $j=1,\ldots,M_p$, and
$-L/2+1\leq n_j \leq -L/4$ or $L/4+1 \leq n_j \leq L/2$ for $j=M_p+1,\ldots,L/2$. Here, $M_p$ and
$L/2-M_p$ are the numbers of Bethe roots with imaginary part zero and $\pi/2$, respectively. We
consider the set $\{\lambda_1,\lambda_2,\lambda_3,\lambda_4\}$ of four possible Bethe roots, corresponding to
the numbers $n$, $L/2+1-n$, $-L/2+n$, and $-n+1$ with $1\leq n \leq L/4$. For small $\phi$ we can
expand them in the small parameter $\delta = \frac{\phi}{2L\cos(\frac{\pi}{L}(2n-1))}$. Setting
$\lambda=\text{artanh}\left(\tan\left[\tfrac{\pi}{L}(n-\tfrac{1}{2})\right]\right)$ we obtain 
\begin{equation}
\begin{array}{l@{\qquad}l}
	\lambda_3 = -\lambda + \tfrac{\i\pi}{2} +\delta + \epsilon \epc & \lambda_2 = \lambda + \tfrac{\i\pi}{2} + \delta + \epsilon \epc \\[1ex]
	\lambda_4 = -\lambda - \delta + \epsilon \epc & \lambda_1 = \lambda - \delta + \epsilon \epp
\end{array}
\end{equation}
These four approximate solutions can be considered as a deformation of the rectangle with corners $(\lambda, \lambda+\i\pi/2, -\lambda+\i\pi/2, -\lambda)$. 

\begin{figure}[tbph]
\centering
\includegraphics[width=0.6\textwidth]{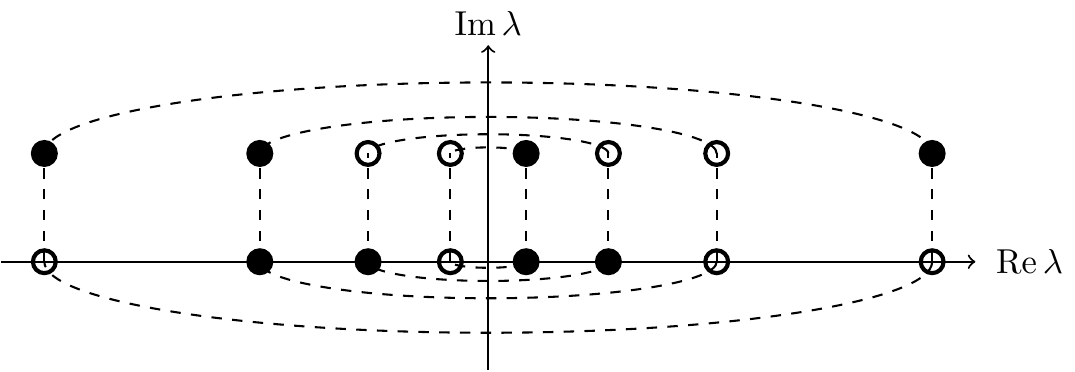}
\caption{Example of a Bethe root configuration (full circles) for a chain of length $L=16$. The four
  `rectangles' are indicated by the dashed lines. The `rectangle condition' is fulfilled, leading to a
  nonzero overlap with the N\'eel state.  Each collection of four possible roots with equal
  $\left|\text{Re}{\lambda}\right|$ contains exactly two roots sitting on two adjacent corners of
  the rectangle.}
 \label{fig:rectangles}
\end{figure}

In the formulas for the overlaps of Ref.~\cite{2014_Brockmann_JPA_47_345003} ---valid also for
the so-called (unnormalized) off-shell Bethe states--- we first send $\epsilon\to 0$ and afterwards
$\phi\to 0$. It can be shown that overlaps of the N\'eel state with normalized Bethe states identified 
by a set of Bethe roots $\{\lambda_j\}_{j=1}^{L/2}$ containing an opposite pair of
the form $(\lambda_1, \lambda_3)$ and/or $(\lambda_2, \lambda_4)$ are of order $\epsilon$. Thus, if
there are three or four Bethe roots in one rectangle, there is at least one opposite pair and the
overlap is zero in the limit $\epsilon \to 0$. Therefore, to have non-vanishing overlap, there can
be at most two XX Bethe roots in each rectangle. Since there are in total $L/2$ Bethe roots and
$L/4$ rectangles, all rectangles are occupied exactly by two Bethe roots sitting at two adjacent
corners (`rectangle condition'). In Fig.~\ref{fig:rectangles} we show an example for which the
rectangle condition is fulfilled.

The overlap formula for Bethe states \cite{2014_Brockmann_JPA_47_145003} consists of two parts, a
prefactor and a ratio of two determinants. It can be shown that the latter is exactly one in the XX
limit, if the rectangle condition is fulfilled, zero otherwise. The remaining part is the prefactor
that simplifies in the XX limit to
\begin{equation}
  \left| \frac{\langle N_{1,2} | \{\lambda_j\}_{j=1}^{L/2}\rangle}{\sqrt{\langle \{\lambda_j\}_{j=1}^{L/2} | \{\lambda_j\}_{j=1}^{L/2} \rangle}}\right|_{\gamma=\pi/2, \phi\to 0} = \prod_{j=1}^{L/4}\frac{\sqrt{\tanh(\mu_j+\frac{\i\pi}{4})\tanh(\mu_j-\frac{\i\pi}{4})}}{2\sinh(2\mu_j)} = 2^{-L/4} \prod_{j=1}^{L/4}\cot\left[\tfrac{2\pi}{L}(n_j-\tfrac{1}{2})\right] \epp
\end{equation}
Here, the Bethe roots $\mu_j$, $j=1,\ldots,L/4$, belong to a solution of the untwisted homogeneous Bethe Ansatz equations, each of them representing one of the $L/4$ rectangles. Due to the $\pi$-periodicity of the tangent function we have $\prod_{n_j=1}^{L/4}\cot^2\left[\frac{2\pi}{L}(n_j-\frac{1}{2})\right] = 1$, and finally obtain
\begin{equation}
  \left| \frac{\langle N_{1,2} | \{\lambda_j\}_{j=1}^M\rangle}{\sqrt{\langle \{\lambda_j\}_{j=1}^M | \{\lambda_j\}_{j=1}^M \rangle}}\right|_{\gamma=\pi/2, \phi\to 0} = \left\{ \begin{array}{cl} 0 & \text{for $M\neq L/2$,} \\ 0 & \text{if there are more (or less) than two roots in one rectangle,} \\ 0 & \text{if there is a rectangle with roots sitting at opposite corners,}\\ 2^{-L/4} & \text{otherwise.} \end{array}\right.
\end{equation}
Here, `otherwise' means that all $L/4$ rectangles are occupied exactly by two roots sitting at adjacent corners, see Fig.~\ref{fig:rectangles} for an example.

The rectangle condition allows for four possibilities for each rectangle and hence, in total, for $4^{L/4}=2^{L/2}$ many states, each contributing to the sum of absolute squares of overlaps with a constant $2^{-L/2}$. This can be translated into the language of momenta of XX Bethe states. The momentum assigned to a Bethe root, which can be interpreted as a single particle momentum, is given by
\begin{equation}
  p_j = p(\lambda_j)\qquad \text{with}\quad p(\lambda) = -\i\ln\left[\frac{\sinh(\lambda+\i\pi/4)}{\sinh(\lambda-\i\pi/4)}\right]\epp
\label{eq:momentum}
\end{equation}
The rule that eigenstates with Bethe roots in opposite corners are forbidden translates into the
rule that the corresponding momenta cannot differ by $\pm\pi$, which is exactly the selection rule
\eqref{eq:selectionrule1}.

\subsection{Application: the Loschmidt echo}\label{sec:LoschEch}

In this section we use our findings for the overlap distribution in the quench
$\Delta_i=\infty\to\Delta_f=0$ to compute the Loschmidt echo (LE) defined in
Eq.~\eqref{eq:LoschEch}. We start from the definition of the Loschmidt echo and exploit the selection rules
of Sec.~\ref{sec:selection} to obtain an explicit product formula. For the sake of simplicity we
only discuss open chains, but the periodic case is very similar. We start by considering the case
with only one of the N\'eel states $\ket{N_1}$ or $\ket{N_2}$, before deriving the result for the
initial state $\ket{\Psi_\sigma} = (\ket{N_1}+\sigma \ket{N_2})/\sqrt{2}$. Recall that the LE is
defined as
\begin{equation}
  \mathcal{L}(t) = |\langle N_{1,2}|\e^{-\i H t}|N_{1,2}\rangle|^2 \epc
\end{equation}
where the Hamiltonian $H$ of the time evolution operator is given in this subsection by the open
chain Hamiltonian of the free system. Expanding $|N_{1,2}\rangle$ in a basis of energy eigenstates
$|n\rangle$, we obtain
\begin{equation}\label{eq:LoschEch1bis}
  \braket{N_{1,2}|\e^{-\i H t}|N_{1,2}}=\sum_{n}\e^{-\i E_n t}|c_n|^2 \epc
\end{equation}
where $E_n$ is the many-body energy and $c_n=\langle N_{1,2}|n\rangle$. As in Sec.~\ref{sec:selection}, we use the sets of single particle momenta $\{q_\a\}_{\a=1}^{L/2}$ to label the eigenstates $|n\rangle$. 

From Secs.~\ref{sec:selection} and \ref{sec:ba} we know that the overlap distribution is flat and that if we take $|N_{1}\rangle$ or $|N_{2}\rangle$ as initial state, only selection rule \ref{rule1} of Sec.~\ref{sec:selection} must be fulfilled, namely $q_\alpha+q_\beta\neq\pi$ for all indices $\alpha$ and $\beta$. In terms of the single particle energies $\epsilon_q = \cos(q)$, the previous condition for the momenta can be interpreted in the following way:
\begin{equation}
  \epsilon_q \neq -\epsilon_{q'} \qquad\textnormal{for all}\quad q,q'\in \{q_\a\}_{\a=1}^{L/2} \epp
\label{eq:enercond}
\end{equation}
Since the total energy of the state is $E(\{q_\a\}_{\a=1}^{L/2}) = \sum_{\a=1}^{L/2}\epsilon_{q_\a}$, Eq.~\eqref{eq:LoschEch1bis} can be rewritten as
\begin{equation}\label{eq:LoscEch2}
  \braket{N_{1,2}|\e^{-\i H t}|N_{1,2}} = 2^{-L/2}{\sum}^\prime \prod_{\a=1}^{L/2}\e^{-\i \epsilon_{q_\a} t} \epc
\end{equation}
where ${\sum}^\prime$ is a sum over all possible states $|\{q_\a\}_{\a=1}^{L/2}\rangle$, i.e., a sum over all subsets $\{q_\a\}_{\a=1}^{L/2}\subset Q$, restricted to the states/subsets for which Eq.~\eqref{eq:enercond} holds. Since half of the possible single particle energies are positive,
 while the others are their negative counterparts, we can rewrite Eq.~\eqref{eq:LoscEch2} as
\begin{equation}
  \braket{N_{1,2}|\e^{-\i H t}|N_{1,2}} = \prod_{\epsilon_q>0}\left(\frac{\e^{-\i\epsilon_q t}+\e^{\i\epsilon_q t}}{2}\right) = \prod_{\epsilon_q>0}\cos\left(\epsilon_q t\right) \epp
\end{equation}
In this way condition \eqref{eq:enercond} is automatically satisfied, since it is impossible, when
expanding the product, to have in a single term the same energies with opposite signs. Therefore,
for the open chain the LE corresponding to one of the N\'eel states $\ket{N_{1,2}}$ is given by the
simple formula 
\begin{equation}
  \mathcal{L}(t) = \prod_{m=1}^{L/2} \cos^2\left[t\cos\left(\frac{\pi m}{L+1}\right)\right] \epp
\label{eq:LoschEch3}
\end{equation}
We note that similar formulas have been derived for periodic chains in
Refs.~\cite{2006_Quan_PRL_96_140604, 2014_Andraschko_PRB_89_125120}. Our result for the open chain
can be also obtained by expressing the LE as a determinant and then computing it explicitly. This is
however unnecessary. As we have seen it can be obtained from simple counting arguments based on the
selection rules of Sec.~\ref{sec:selection}. 

\begin{figure}[tbph]
\centering
\includegraphics[width=0.82\textwidth]{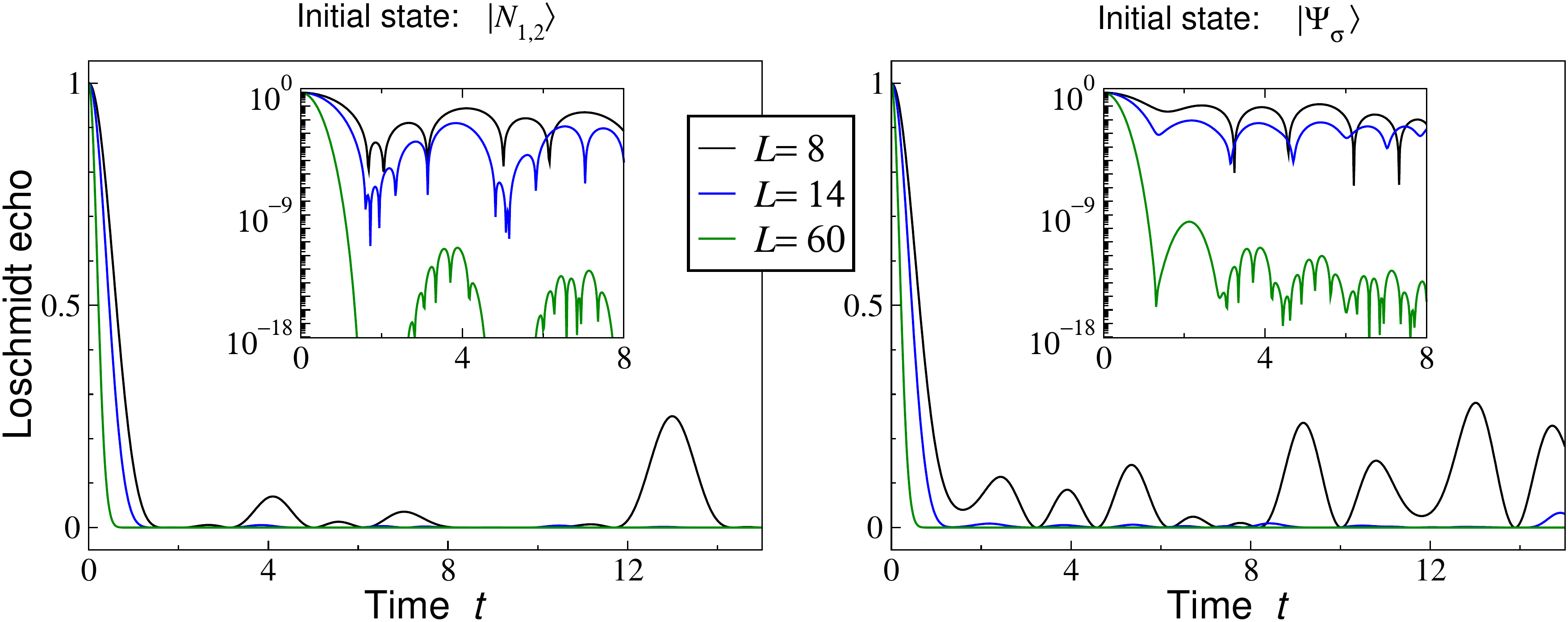}
\caption{
  Loschimdt echoes in open chains of sizes $L=8,14,60$ corresponding to the N\'eel initial
  state $\ket{N_{1,2}}$ (left), see Eq.~\eqref{eq:LoschEch3} and to the initial state
  $\ket{\Psi_\sigma}$ (right), see Eq.~\eqref{eq:finalle}. For the two smaller sizes, a comparison
  with Loschmidt echoes numerically calculated by means of exact diagonalization shows exact agreement with the derived
  closed formulas.}
\label{fig:LoschEch}
\end{figure}

We now consider $\ket{\Psi_\sigma}$ as initial state. The LE is still given by a formula similar to
Eq.~\eqref{eq:LoscEch2},
\begin{equation}\label{eq:LoscEch2bis}
  \braket{\Psi_\sigma|\e^{-\i H t}|\Psi_\sigma} = 2^{-L/2}{\sum}^\prime \prod_{\a=1}^{L/2}\e^{-\i \epsilon_{q_\a} t}\epc
\end{equation}
where the prime at the sum now means that both selection rules \ref{rule1} and \ref{rule2} of
Sec.~\ref{sec:selection} must be fulfilled. In addition to Eq.~\eqref{eq:enercond}, the set of
pseudo-momenta have to satisfy $(-1)^n=\sigma=(-1)^{L/2}$, where $n$ is the number of pseudo-momenta
with negative energy. Equation \eqref{eq:LoscEch2bis} can then be recast as
\begin{equation}\label{LoscEch3}
  \braket{\Psi_\sigma|\e^{-\i H t}|\Psi_\sigma} =\prod_{\epsilon_q>0}\left(\frac{\e^{-\i\epsilon_q t}+\e^{\i\epsilon_q t}}{2}\right) \;+\; \sigma \prod_{\epsilon_q>0}\left(\frac{\e^{-\i\epsilon_q t}-\e^{\i\epsilon_q t}}{2}\right) = \prod_{\epsilon_q>0} \cos(\epsilon_q t) \;+\;  \,\i^{L/2}\prod_{\epsilon_q>0} \sin(\epsilon_q t)\epp
\end{equation}
A simple way to prove this is to expand the two products in Eq.~\eqref{LoscEch3}, and check that all states that do not satisfy the selection rules \ref{rule1} and \ref{rule2} cancel. Therefore, the LE corresponding to $\ket{\Psi_\sigma}$ is given by
\begin{equation}\label{eq:finalle}
  {\cal L}(t)=\left|\braket{\Psi_\sigma|\e^{-\i H t}|\Psi_\sigma}\right|^2 = \left|
  \prod_{m=1}^{L/2} \cos\left[t\cos\left(\frac{\pi m}{L+1}\right)\right] \;+\; 
\i^{L/2}\prod_{m=1}^{L/2}  \sin^2\left[t\cos\left(\frac{\pi m}{L+1}\right)\right] \right|^2
\end{equation}
Both results \eqref{eq:LoschEch3} and \eqref{eq:finalle} are plotted in Fig.~\ref{fig:LoschEch} for
several system sizes $L$. 

\subsection{Application: time-dependent two-point correlation functions}\label{sec:correlators}

We now provide a second application of the results on overlap distributions in
Sec.~\ref{sec:selection} for the quench $\Delta_i=\infty\to\Delta_f=0$.  We will compute the full
time-dependence of finite distance two-point correlation functions after such a quench.  For
simplicity we will restrict to the periodic chain and list results which are derived in the
large-$L$ limit.  We provide only a skeleton outline of the derivation, pointing out how the
selection rules are used.

The time evolution occurs under the $\Delta = 0$ Hamiltonian $H_{\text{xx}}$.  The equal-time
two-point correlators are defined as
\begin{align}
  G_{\Psi_0}^{zz}(j,j+n;t) &= \langle \Psi_0| \e^{\i t H_{\textnormal{xx}}} S_j^{z}S_{j+n}^{z} \e^{-\i t H_{\textnormal{xx}}} |\Psi_0\rangle\epc \\
  G_{\Psi_0}^{+-}(j,j+n;t) &= \langle \Psi_0| \e^{\i t H_{\textnormal{xx}}} S_j^{+}S_{j+n}^{-} \e^{-\i t H_{\textnormal{xx}}} |\Psi_0\rangle\epp
\end{align}
We refer to them as the longitudinal and transverse two-point correlators, respectively. We consider
as initial state $|\Psi_0\rangle$ the four different cases $|\Psi_{+}\rangle =
(|N_1\rangle+|N_2\rangle)/\sqrt{2}$, $|\Psi_{-}\rangle = (|N_1\rangle -|N_2\rangle)/\sqrt{2}$ and
the two N\'eel states $|N_1\rangle$ and $|N_2\rangle$ themselves. Since the state
$|\Psi_{\sigma}\rangle$ is an eigenstate of the translation operator $\hat{T}$, the two-point
correlators only depend on the distance between the two points, $G_{\Psi_\sigma}^{zz,+-}(j,j+n;t) =
G_{\Psi_\sigma}^{zz,+-}(n;t) = \frac{1}{L}\sum_{j=1}^L G_{\Psi_\sigma}^{zz,+-}(j,j+n;t)$.
Applying the Jordan Wigner mapping \eqref{eq:JW} we obtain
\begin{align}\label{eq:longit_corr}
  G_{\Psi_0}^{zz}(n;t) &= \frac{1}{L}\sum_{j=1}^L \langle \sigma_j^{z}\sigma_{j+n}^z \rangle = \frac{1}{L}\sum_{j=1}^L \langle \left(c_j^{\dag}c_j^{\phantom\dag}-\frac{1}{2}\right)\left(c_{j+n}^{\dag}c_{j+n}^{\phantom\dag}-\frac{1}{2}\right) \rangle = \frac{1}{L}\sum_{j=1}^L \langle c_j^{\dag}c_j^{\phantom\dag}c_{j+n}^{\dag}c_{j+n}^{\phantom\dag}\rangle -\frac{1}{4}\\
  G_{\Psi_0}^{+-}(n;t) &= \frac{1}{L}\sum_{j=1}^L \langle c_j^{\dag} \left[\prod_{m=1}^{n-1}(1-2c_{j+m}^{\dag}c_{j+m}^{\phantom\dag})\right] c_{j+n}^{\phantom\dag}\rangle\epc
\end{align}
where from now on the time dependence is implicit, i.e., the expectation values are taken between
the time evolved states, $\braket{\cdot}=\braket{\Psi(t)|\cdot|\Psi(t)}$ with $\ket{\Psi(t)}=\e^{-\i
  Ht}\ket{\Psi_0}$.  In the first line we used that the initial state $|\Psi_0\rangle$ lies in the zero
magnetization sector and therefore $\frac{1}{L} \sum_{j=1}^L \langle
c_j^{\dag}c_j^{\phantom\dag}\rangle = 1/2$ for all times $t$. In the second line we wrote
the `Jordan-Wigner string' operators as $\e^{\i\pi c_j^{\dag}c_j^{\phantom\dag}} =
1-2c_j^{\dag}c_j^{\phantom\dag}$, which can be written as
\begin{equation}
  1 - 2c_j^{\dag}c_j^{\phantom\dag} = 1 - \frac{2}{L}\sum_{p} \tilde{c}_p^{\dag}\tilde{c}_p^{\phantom\dag} - \frac{2}{L}\sum_{\substack{p,q\\p\neq q}} \e^{-\i(p-q)j}\tilde{c}_p^{\dag}\tilde{c}_q^{\phantom\dag}\epp
\end{equation}
It turns out that the first two terms cancel up to order $1/L$, so that the last term dominates in
the large $L$ limit.  Thus, the transverse correlator will involve correlators of products like
$\left[\prod_{m=1}^{n}c_{p_m}^{\dag}\right]\left[\prod_{m=1}^{n}c_{q_m}^{\phantom\dag}\right]$,
where $\{p_m\}$ and $\{q_m\}$ are sets of $n$ distinct single-particle momenta.  As an example, the
correlator $G_{\Psi_\sigma}^{+-}(n;t)$ is a momentum sum over products of phase factors, which can be 
rewritten as  
\begin{align}
  \langle \Psi_\sigma|
  \left[\prod_{m=1}^{n}c_{p_m}^{\dag}\right]\left[\prod_{m=1}^{n}c_{p_{A(m)+\pi}}^{\phantom\dag}\right]|\Psi_\sigma\rangle
  &= \sum_{\{k\},\{k'\}}\langle \Psi_\sigma|\{k\}\rangle\langle \{k'\}|\Psi_\sigma\rangle\;\;
  \langle
  \{k'\}|\left[\prod_{m=1}^{n}c_{p_m}^{\dag}\right]\left[\prod_{m=1}^{n}c_{q_{m}}^{\phantom\dag}\right]|\{k\}\rangle
  \notag\\ 
  &= (-1)^{\frac{n(n-1)}{2}} \left(\frac{1+(-1)^n}{2}\right) \frac{(-1)^{|A|}}{2^n}\epc
\label{eq:matrix_elements_transverse}
\end{align}
The second line has been obtained using the selection rules of Sec.~\ref{sec:selection}.  Since $n$
momenta of the eigenstates $|\{k_\a\}_{\a=1}^{L/2}\rangle$ that obey condition
\eqref{eq:selectionrule1} are fixed, there are $\frac{1}{2}\sum_{j=0}^{L/2-n} \binom{L/2-n}{j} =
2^{L/2-n-1}$ states with overlap square $2^{-L/2+1}$. This gives the factor $2^{-n}$. The factor
$(1+(-1)^n)/2$ comes from the additional selection rule (`number of pair flips') for the total
momentum of the eigenstates $|\{k\}\rangle$ and $|\{k'\}\rangle$. Hence, the transverse two-point
correlator at points $j$ and $j+n$ is only nonzero for $n$ even.  Here, $A$ is the
mapping matrix from the set $\{p_m\}$ to the set $\{q_m\}$, given by an element of the symmetric group
$\mathcal{S}^n$. The factor $(-1)^{|A|}$ implies that the result is a determinant.  Carrying all
factors and simplifying, we eventually obtain ($t\geq 0$)
\begin{align}
  \hspace{4cm} G_{\Psi_\sigma}^{+-}(n;t) &=  \frac{1}{2} \det\nolimits_{n}\Big(\tilde{J}_{m-m'-1}(2t)\Big) && \text{for $n$ even} \epc\hspace{4cm} 
\label{eq:transverse_correlator_even}\\
  \hspace{4cm} G_{\Psi_\sigma}^{+-}(n;t) &= 0 &&\text{for $n$ odd}\epp\hspace{4cm} \label{eq:transverse_correlator_odd}
\end{align}
Here, we have defined the quantity
\begin{equation}\label{eq:Bessel_finite_L}
  \tilde{J}_{m}(2t) = \frac{(-\i)^m}{L}\sum_{j=1}^L \e^{2\i t \cos(p_j)-\i m p_j}\epp
\end{equation}
Similarly, one obtains for the N\'eel states themselves
\begin{equation}\label{eq:transverse_correlator_Neel}
  G_{N_a}^{+-}(j,j+n;t) = \frac{(-1)^{n(j+a-1)}}{2\i}\det\nolimits_{n}\Big(\tilde{J}_{m-m'-1}(2t)\Big)\epp
\end{equation}
In the derivation of the formulas for $G_{\Psi_0}^{+-}$ some $1/L$ terms were dropped.  However, we
have found that formulas \eqref{eq:transverse_correlator_even} and \eqref{eq:transverse_correlator_odd}
for $G_{\Psi_\sigma}^{+-}(n;t)$ are exact for $n=2,4$ and for any finite even $L$.

The longitudinal correlators are simpler to derive, since they have no Jordan-Wigner strings. The result is 
\begin{equation}
  G_{\Psi_0}^{zz}(n;t) = \frac{1}{4}\Big(\delta_{n,0}-\tilde{J}_n^2(2t)+(-1)^n \tilde{J}_0^2(2t)\Big)
  \label{eq:longitudinal_correlator}
\end{equation}
for all four initial states $|\Psi_0\rangle = \ket{\Psi_{\pm}}, \ket{N_{1,2}}$.

In the thermodynamic limit, the summation in Eq.~\eqref{eq:Bessel_finite_L}
becomes the integral that defines the $m$-th order Bessel function,
\begin{equation}
\tilde{J}_{m}(2t)\xrightarrow{L\to\infty}J_{m}(2t)\epp
\end{equation}
Thus, the above formulas for $G_{\Psi_0}^{+-}$ and $G_{\Psi_0}^{zz}$ are also valid in the thermodynamic limit, with
the substitution $\tilde{J}_{m}\to J_{m}$.
Previously, Ref.~\cite{2010_Barmettler_NewJPhys_12_055017} presented (connected versions of)
longitudinal correlators in the thermodynamic limit.  To the best of our knowledge, the determinant
formulas for the more involved transverse correlators have not appeared previously.

\section{Quenches from \texorpdfstring{$\Delta_{i}=\infty$}{} to  \texorpdfstring{$\Delta_{f}\ll 1$}{}:
  perturbative treatment of the splitting  }\label{sec:A}
  
Quenches from the initial state $\ket{\Psi_\sigma}$ ($\Delta_i=\infty$) to small but nonzero values
of $\Delta_f$ display a `splitting' of the overlap distribution, as discussed in
Sec.~\ref{sec:numeric_infty2various} and shown in Figs.~\ref{fig:smalldelta}(a,b).  In this section,
we will use first order perturbation theory in $\Delta_f$ to characterize this splitting.

\subsection{First order perturbation expansion} 

For the final Hamiltonian $H = H_{\textnormal{xx}} + \Delta_f H_{\textnormal{z}}$, we consider
$\Delta_f H_{\textnormal{z}}$ as a perturbative potential $V$.  Unfortunately, the unperturbed
Hamiltonian $H_{\textnormal{xx}}$ is highly degenerate.  In particular, among the states which obey
the selection rules, there are many which are degenerate.  Hence, a full calculation requires the
use of degenerate perturbation theory for many of the eigenstates.  However, our purpose here is to
provide a basic understanding of the splitting and the general form of the overlap distribution for
small nonzero $\Delta_f$.  Hence, it will suffice to concentrate on those states which have
non-degenerate energy eigenvalues, and use the much simpler non-degenerate perturbation theory.  We
will show that this treatment recovers the main features observed numerically.

For the sake of simplicity we consider a periodic chain with $L/2$ even, which means that in the
zero magnetization sector there is an even number $N=L/2$ of fermions in the system. All the
calculations can be trivially extended to the odd particle case. In terms of the Jordan-Wigner
fermions of Eq.\ \eqref{eq:JW}, the perturbative potential $V=\Delta_f H_{\textnormal{z}}$ can be
written as 
\begin{equation}
  V = \Delta_f\sum_{j=1}^{L} \left(n_j-\frac{1}{2}\right)\left(n_{j+1}-\frac{1}{2}\right) = \Delta_f \sum_{j=1}^{L} \left( n_j n_{j+1}-\frac{1}{2}(n_j+n_{j+1})+\frac{1}{4}\right) \epc
  \label{eq:pertpot}
\end{equation} 
where $n_j=c_j^\dag c_j^{\phantom\dag}$. We can drop the constant term and the terms proportional to the 
particle number $\hat{N} = \sum_{j=1}^L n_j$ since they represent just shifts of the energy and are 
irrelevant for a perturbative expansion. Using again momentum operators defined in Eq.~\eqref{eq:Fourier} 
the only relevant contribution can be written as
\begin{equation}
  \sum_{j=1}^L n_j n_{j+1} = \frac{1}{L^2}\sum_{j=1}^{L}\sum_{p,q,r,s} \e^{-\i(p-q+r-s)j}\e^{-\i(r-s)} \tilde{c}_p^\dag\tilde{c}_q^{\phantom\dag}\tilde{c}_r^\dag\tilde{c}_s^{\phantom\dag} \epp
\end{equation}
Carrying out the sum over the spatial indices we find the condition $q+s=p+r$, which is nothing but the 
momentum conservation. After these considerations we can rewrite the perturbative potential for the 
periodic chain as
\begin{equation}
  V = \frac{\Delta_f}{L}\sum_{p,q,r} \e^{\i(p-q)} \tilde{c}_p^\dag \tilde{c}_q^{\phantom\dag} \tilde{c}_r^\dag \tilde{c}_{p-q+r}^{\phantom\dag} \epp
  \label{eq:pertur}
\end{equation}
Let us now formally write the expression for the corrected overlap. Given $|\Phi\rangle$ the
`corrected eigenstate' with $\lim_{\Delta_f\to 0} |\Phi\rangle = |\{k_\a\}_{\a=1}^{L/2}\rangle$, the
latter being an eigenstate of the XX chain that obeys the selection rules (`pair condition'
\eqref{eq:selectionrule1} and zero total momentum), we obtain
\begin{equation}
  \langle\Psi_\sigma|\Phi\rangle = \sqrt{2}\left(\frac{1}{L}\right)^{L/4}\Big(\det_{1\leq j,\a \leq L/2}\left(\e^{2\i k_\a j}\Big)+\sum_{|\{k'\}\rangle}\frac{\langle\{k\}|V|\{k'\}\rangle}{E_k-E_{k'}}\det_{1\leq j,\a \leq L/2}\Big(\e^{2\i k'_\a j}\Big)\right) \epc
  \label{eq:overcorr}
\end{equation}
where the states $|\{k\}\rangle = |\{k_\a\}_{\a=1}^{L/2}\rangle$ and $|\{k'\}\rangle = |\{k'_\a\}_{\a=1}^{L/2}\rangle$ 
are eigenstates of the free fermionic Hamiltonian. Here, we only choose states $|\{k\}\rangle$ for which the 
energy value $E=\sum_{\a=1}^{L/2} \cos(k_\a)$ is non-degenerate, such that the energy difference in 
the denominator is never zero. Normalizing with $\langle\Psi_\sigma|\{k\}\rangle$ we obtain
\begin{equation}
  \left|\frac{\langle\Psi_\sigma|\Phi\rangle}{\langle\Psi_\sigma|\{k\}\rangle}\right| = 1-{\sum_{|\{k'\}\rangle}}^\prime\frac{|\langle\{k\}|V|\{k'\}\rangle|}{E_k-E_{k'}} \epc
  \label{eq:overcorrnorm}
\end{equation}
where we used that the determinant under the sum is either zero or always given by the same absolute value 
$|\det_{1\leq j,\a \leq L/2}(\e^{2\i k_\a j})|$. The prime at the sum means that it is restricted to the 
`nonzero overlap states'. The perturbative potential \eqref{eq:pertur} connects states $|\{k\}\rangle$ 
and $|\{k'\}\rangle$ in which just two momenta are exchanged. Due to the selection rule 
\eqref{eq:selectionrule1} we have that
\begin{equation}
  |k_\alpha-k'_\alpha|=\pi\epc  \qquad|k_\beta-k'_\beta|=\pi\epp
\end{equation}
Therefore, the energy difference in the denominator can be written as
\begin{equation}
  E_k-E_{k'} = \cos(k_\alpha) + \cos(k_\beta) - \cos(k'_\alpha)-\cos(k'_\beta) = 2\left(\cos(k_\alpha) + \cos(k_\beta)\right) \epp
\end{equation}
Furthermore, for the numerator, we obtain after a short calculation 
\begin{equation}
  |\langle\{k\}|V|\{k'\}\rangle| = \frac{4\Delta_f}{L}\sin^2\left(\frac{k_\alpha-k_\beta}{2}\right) \epp
\label{eq:potential_explicit}
\end{equation}
The sign of $\langle\{k\}|V|\{k'\}\rangle$ is fixed by the order of momenta of the states
$|\{k\}\rangle$ and $|\{k'\}\rangle$. It exactly cancels the sign coming from the determinant that
we pulled out of the sum in Eq.~\eqref{eq:overcorr}. The relative minus sign in front of the sum in
Eq.~\eqref{eq:overcorrnorm} is due to the factor $\e^{\i(p-q)}$ in Eq.~\eqref{eq:pertur} and comes
from the fact that the momenta $k_\alpha$, $k_\beta$ and $k'_\alpha$, $k'_\beta$ differ by
$\pm\pi$. In summary, we find 
\begin{equation}
  \frac{\left|\langle\Psi_\sigma|\Phi\rangle\right| - \left|\langle\Psi_\sigma|\{k\}\rangle\right|}{\left|\langle\Psi_\sigma|\{k\}\rangle\right|} = -\frac{2\Delta_f}{L}\sum_{\substack{\alpha,\beta=1\\ \alpha<\beta}}^{L/2} \frac{\sin^2\left(\frac{k_\alpha-k_\beta}{2}\right)}{\cos(k_\alpha) + \cos(k_\beta)} \epp
  \label{eq:overcorrnorm2}
\end{equation}
Follwoing the definition \eqref{eq:normalized_shift_defn} in Sec.~\ref{sec:numeric_infty2various},
we call the quantity in Eq.~\eqref{eq:overcorrnorm2} the ``normalized shift'' and the numerator of
the left-hand side just the ``shift''.

\begin{figure}[tbph]
\centering
\includegraphics[width=0.65\textwidth]{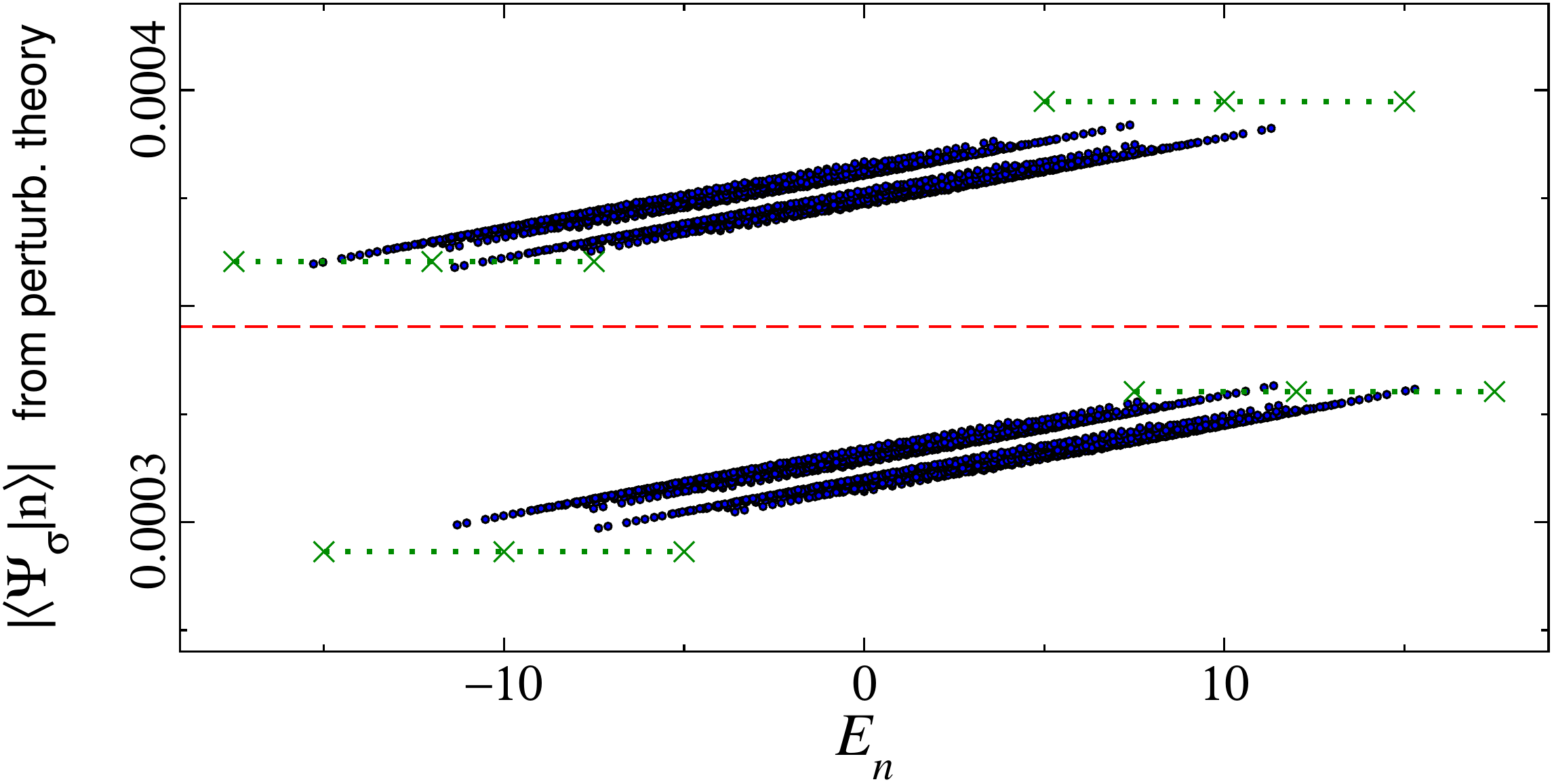}
\caption{Overlap magnitudes for a quench from $\Delta_i=\infty$ to small $\Delta_f$, obtained from
  first-order non-degenerate perturbation theory. The specific values plotted here are for
  $\Delta_f=0.01$ and $L=48$.  Only non-degenerate states are included. The dashed line depicts the
  unperturbed ($\Delta_f=0$) value $2^{-L/4+1/2}$. The four horizontal lines (dots and crosses)
  indicate the extremal values of the shift acoording to Eqs.~\eqref{eq:TDL_double_sum} and
  \eqref{eq:largest_split}.}
  \label{fig:perturbative_shift}
\end{figure}

In Fig.~\ref{fig:perturbative_shift} we plot the perturbative results \eqref{eq:overcorrnorm2} for
the nonzero overlaps for $\Delta_f=0.01$ and $L=48$.  (We plot the predictions for actual overlaps
instead of the normalized shifts.)  The perturbative treatment only covers non-degenerate states.
Hence, a quantitative comparison with the numerical data of Fig.~\ref{fig:smalldelta}(a) is not
appropriate.  However, non-degenerate perturbation theory clearly reproduces the main features of
the data.  Figure \ref{fig:perturbative_shift} shows the overall shape to be very similar to that
seen in the inset to Fig.~\ref{fig:smalldelta}(a), with two branches each having an overall upward
slope.  The perturbative approach explains the shift being proportional to $\Delta_f$ (by providing
a non-vanishing contribution at first order in $\Delta_f$, see above).  In addition, we will show in the
subsection below that the result \eqref{eq:overcorrnorm2} implies that the smallest and largest
normalized shifts grow linearly with $L$, so that the normalized average branch shifts grow linearly with $L$,
as found numerically, see Fig.~\ref{fig:smalldelta}(b). 

The perturbative prediction in Fig.~\ref{fig:perturbative_shift} also shows some substructures,
e.g., each branch is futher split (recursively) into sub-branches.  It is unclear whether such
substructures might survive when the degenerate states are taken into account, or when higher-order
corrections are included.  In addition, substructures in the XXZ spectrum are often heavily
determined by boundary conditions \cite{2010_Haque_PRA_82_012108, 2014_Sharma_PRA_89_043608,
  2013_Alba_2013_JStatMech_P10018}. So, they can be expected to be different for periodic and open
boundary conditions.  A detailed study of such substructures is beyond the scope of the current
work.

\subsection{ \texorpdfstring{$L$}{}-dependence of smallest and largest shifts}

In order to see how the normalized shift scales with $L$ we compute the sum in
Eq.~\eqref{eq:overcorrnorm2} for large system sizes $L$. First of all, we observe that the shift is
minimal in absolute value and opposite for the ground state and the highest energy state. For the
latter the momenta are uniformly distributed between zero and $\pi/2$ as well as between $3\pi/2$
and $2\pi$ or equivalently, due to the periodicity of the trigonometric functions, between $-\pi/2$
and $\pi/2$. So, $k_\a=\frac{2\pi}{L} (\a-1/2)$ with $-L/4+1 \leq \a \leq L/4$ and the sum over
momenta turns in the thermodynamic limit into an integral from $-\pi/2$ to $\pi/2$ with prefactor
$L/(2\pi)$. Therefore, the double sum of the right-hand side of Eq.~\eqref{eq:overcorrnorm2} can be
written as
\begin{equation}
  \sum_{\substack{\alpha,\beta=1\\ \alpha<\beta}}^{L/2}
  \frac{\sin^2\left(\frac{k_\alpha-k_\beta}{2}\right)}{\cos(k_\alpha) + \cos(k_\beta)}
  \;\longrightarrow\; \frac{1}{2}\left(\frac{L}{2\pi}\right)^2
  \int_{-\pi/2}^{\pi/2}\int_{-\pi/2}^{\pi/2} \frac{\sin^2\left(\frac{k-q}{2}\right)}{\cos(k) +
    \cos(q)}dk dq = \frac{L^2}{2}\left(\frac{1}{4}-\frac{1}{2\pi}\right)\epp 
\label{eq:TDL_double_sum}
\end{equation}
Together with the small prefactor $2\Delta_f/L$ in the perturbative potential \eqref{eq:pertur} 
this eventually leads to a normalized shift of $-L \Delta_f(1/4-1/(2\pi))\approx -0.09085\,L\Delta_f $ 
for the highest energy state and, similarly, $L \Delta_f(1/4-1/(2\pi)) \approx 0.09085\,L\Delta_f $ for 
the ground state. This means that as we increase the system size $L$ we have to consider smaller and 
smaller values of $\Delta_f$, i.e.~$\Delta_f \ll 1/L$. Otherwise, we are not in the perturbative 
regime anymore. To summarize, if we increase the value of the final anisotropy from $\Delta_f=0$ to 
$\Delta_f \ll 1/L$, a gap opens in the nontrivial part of the operlap distribution, which we call 
``splitting'' and which scales as $2^{-L/4+1/2}L$ with prefactor $\Delta_f(1/2-1/\pi)$, i.e.~twice 
the scaling factor of the shift of the ground state.

Furthermore, the largest value of the shift can be easily computed, too. Our numerical investigation 
shows that the state that belongs to the largest split is basically given by the set of momenta of 
the highest energy state where the two innermost momenta $\pm \pi/L$ are replaced by the two momenta $\pm (\pi - \pi/L)$. This leads to an additional contribution with respect to the dominating part of the highest energy state, i.e.~$L\Delta_f(1/(2\pi)-1/4)$, of
\begin{align}
  -\frac{2\Delta_f}{L}\cdot 2\sum_{k_\beta}\frac{\sin^2\left(\frac{\pi}{2}-\frac{\pi}{2L}-\frac{k_\beta}{2} \right)}{\cos(\pi-\pi/L)+\cos(k_\beta)} &= -\frac{2\Delta_f}{L}\cdot 4 \sum_{m=2}^{L/4}\frac{\sin^2\left(\frac{\pi}{2}-\frac{\pi}{2L}-\frac{\pi}{2L}(2m-1) \right)}{-\cos(\pi/L)+\cos(\frac{\pi}{2L}(2m-1))} \notag\\
  &\approx \frac{8\Delta_f}{L}\sum_{m=2}^{L/4}\frac{1}{1-\frac{\pi^2}{2L^2}-(1-\frac{\pi^2}{2L^2}(2m-1)^2)} = \frac{4L\Delta_f}{\pi^2}\sum_{m=2}^{L/4}\frac{1}{m(m-1)} \approx \frac{4L\Delta_f}{\pi^2}\epp
\label{eq:largest_split}
\end{align}
The first factor $2$ in front of the sum comes from the fact that both innermost momenta are replaced,
the second factor $2$ from the fact that the remaining momenta $k_\beta$ are symmetrically distributed
around zero (modulo periodicity of $2\pi$). Hence, the largest shift also scales like
$2^{-L/4+1/2}L$, now with the prefactor $\Delta_f(4/\pi^2+1/(2\pi)-1/4) \approx
0.31444\,\Delta_f$. 

In Fig.~\ref{fig:perturbative_shift}, the large-$L$ predictions for maximum and minimum values of
the shifted overlaps, Eqs.~\eqref{eq:TDL_double_sum} and \eqref{eq:largest_split}, are indicated 
with horizontal dotted/crossed lines.

\section{Quenches from \texorpdfstring{$\Delta_i\gg1$}{} to  \texorpdfstring{$\Delta_f=0$}{}: analytical results }\label{sec:C}

In this section we use a perturbative approach to explain one of the features of the overlap
distribution that can be observed in the quench protocol from $\Delta_i$ large but finite to the
free fermion point $\Delta_f=0$ . As it can be seen in Fig.~\ref{fig:fromlarge}, the nontrivial part of
the overlap distribution is rather smooth as a function of $E_n$, and decreases roughly linearly
with $E_n$.  The slope increases with decreasing $\Delta_i$.  This behavior is in sharp contrast to
the splitting structures of the $\Delta_i=\infty \to \Delta_f\ll 1$ case.
In the following we will use perturbation theory in $1/\Delta_i$ to show that the normalized shift
is proportional to $E_n$ at first order (for periodic chains), which explains the smooth behavior of
the $|c_n|$'s as a function of $E_n$.  We will also compute the quadratic correction by making use
of second order perturbation theory.

For simplicity we consider a periodic chain. In the case of an open chain the main steps of the
following calculation can be similarly performed, but there are complications because of the edges. 
The initial state $\ket{\Psi_0}$, which is the ground state at large but finite $\Delta_i$, is
adiabatically connected to the state  $\ket{\Psi_\sigma}$, which we have used as the initial state
for $\Delta_i=\infty$:
\begin{equation}\label{eq:zeroth_order}
  \lim_{\Delta_i\to\infty} \ket{\Psi_0} = \ket{\Psi_\sigma} \qquad\text{with}\quad \sigma = (-1)^{L/2}\epp
\end{equation}
In order to make the perturbative calculation mathematically stringent, we divide the initial 
Hamiltonian (and henceforth all corresponding energies) by $\Delta_i$. We  consider the second 
term in 
\begin{equation}\label{eq:perturbed_Hamiltonian}
H_i/\Delta_i = \sum_{j=1}^L S_j^z S_{j+1}^z + \frac{1}{\Delta_i} \sum_{j=1}^L \frac{1}{2}\left(S_j^+
  S_{j+1}^- + S_j^- S_{j+1}^+\right) = H_{\textnormal{z}} + \Delta_i^{-1} H_{\text{xx}}
\end{equation}
as a small perturbation. 

In order to analyze the overlaps $\langle \{k_\a\}_{\a=1}^{L/2}|\Psi_0\rangle$, where $
|\{k_\a\}_{\a=1}^{L/2}\rangle$ are the eigenstates of the final Hamiltonian with $\Delta_f=0$,
perturbatively, we expand the ground state $\ket{\Psi_0}$ of $H_i$ in the small parameter
$1/\Delta_i$,  
\begin{equation}\label{eq:pert_expa_state}
  \ket{\Psi_0} = \ket{\Psi_0^{(0)}} +  \Delta_i^{-1} \ket{\Psi_0^{(1)}} +  \Delta_i^{-2} \ket{\Psi_0^{(2)}} + \ldots
\end{equation}
The zeroth-order (unperturbed) wavefunction is $\ket{\Psi_0^{(0)}} = \ket{\Psi_\sigma}$. 

The energy eigenvalue $E_\sigma$ of $\ket{\Psi_\sigma}$ is two-fold degenerate at zeroth order and
the corresponding other state is simply $\ket{\Psi_{-\sigma}}$. Since the two states
$\ket{\Psi_{-\sigma}}$ and $|\Psi_0\rangle$ (with $\sigma$ fixed) are both eigenstates of the translation
operator 
with opposite eigenvalues $-\sigma$ and $\sigma$, we deduce
$\langle\Psi_{-\sigma}|\Psi_0\rangle=0$. This means that $\ket{\Psi_{-\sigma}}$ does not appear in any expansion 
of $\ket{\Psi_{0}}$. We can therefore apply non-degenerate perturbation theory. Using $\ket{\Psi_{0}^{(0)}} = 
\ket{\Psi_{\sigma}}$ the higher order corrections read
\begin{align}\label{eq:first_order}
  \ket{\Psi_{0}^{(1)}} &= \sum_{\ket{n^{(0)}}\neq \ket{\Psi_{\pm\sigma}}} \ket{n^{(0)}} \frac{\bra{n^{(0)}} H_{\text{xx}} \ket{\Psi_{\sigma}}}{E_\sigma - E_n^{(0)}} \epc\\
  \ket{\Psi_{0}^{(2)}} &= \sum_{\ket{m^{(0)}},\ket{n^{(0)}}\neq \ket{\Psi_{\pm\sigma}}} \ket{m^{(0)}} \frac{\bra{m^{(0)}} H_{\text{xx}} \ket{n^{(0)}}}{E_\sigma - E_m^{(0)}}\frac{\bra{n^{(0)}} H_{\text{xx}} \ket{\Psi_{\sigma}}}{E_\sigma - E_n^{(0)}} \notag\\
  &\quad - \sum_{\ket{n^{(0)}}\neq \ket{\Psi_{\pm\sigma}}} \ket{n^{(0)}} \frac{\bra{n^{(0)}} H_{\text{xx}} \ket{\Psi_{\sigma}}\bra{\Psi_{\sigma}} H_{\text{xx}} \ket{\Psi_{\sigma}}}{(E_\sigma - E_n^{(0)})^2} \;-\; \frac{1}{2} \ket{\Psi_{\sigma}} \sum_{\ket{n^{(0)}}\neq \ket{\Psi_{\pm\sigma}}} \frac{|\bra{n^{(0)}} H_{\text{xx}} \ket{\Psi_{\sigma}}|^2}{(E_\sigma - E_n^{(0)})^2} \epc \label{eq:second_order}
\end{align}
where $E_n^{(0)}$ is the eigenvalue of $H_{\text{z}}$ with eigenstate $\ket{n^{(0)}}$. 

We first compute the matrix elements in the numerator of Eq.~\eqref{eq:first_order}:
\begin{equation}\label{eq:numerator_first_order}
  \bra{n^{(0)}} H_{\text{xx}} \ket{\Psi_{\sigma}} = \frac{1}{\sqrt{2}}\bra{n^{(0)}}
  \sum_{j=1}^{L}\frac{1}{2}\left(S_j^+S_{j+1}^- + S_j^-S_{j+1}^+\right) \left(\ket{N_1} + \sigma
    \ket{N_1} \right) = \frac{1}{2\sqrt{2}} \left(\delta_{\ket{n^{(0)}}\in S_1} +
    \sigma \delta_{\ket{n^{(0)}}\in S_2}\right)\epc 
\end{equation}
where $S_{1,2}$ are the spaces spanned by the states $\ket{s_j^{(1,2)}} = (S_j^+S_{j+1}^- +
S_j^-S_{j+1}^+)\ket{N_{1,2}}$, $j=1,\ldots,L$, i.e., the space spanned by the spin configurations
obtained via a single neighboring spin pair flip from a N\'eel state.
The energies $E_n^{(0)}$ are all the same, independent of the choice of $\ket{n^{(0)}}\in S_{1,2}$
and given by $E_n^{(0)} = -L/4+1$. Therefore, $E_\sigma - E_n^{(0)} = -1$ for all $\ket{n^{(0)}}\in
S_{1,2}$ and, henceforth,
\begin{equation}\label{eq:first_order_result}
  \ket{\Psi_0^{(1)}} = -\frac{1}{\sqrt{2}} \left(\frac{1}{2}\sum_{\ket{n^{(0)}}\in S_1}\ket{n^{(0)}} + \sigma\frac{1}{2}\sum_{\ket{n^{(0)}}\in S_2}\ket{n^{(0)}}\right) = -H_{\textnormal{xx}} \ket{\Psi_\sigma} \epp
\end{equation}
Thus, the first-order term is obtained by application of $H_\textnormal{xx}$ on the initial state (a similar trick was used in Ref.~\cite{2014_Luitz_JStatMech_P08007}).

Using that the states $|\{k\}\rangle = |\{k_\a\}_{\a=1}^{L/2}\rangle$ are eigenstates of the 
Hamiltonian $H_{\textnormal{xx}}$ with energy $E_k$ we obtain up to first order
\begin{equation}\label{eq:overlap_ratio_first_order}
  \left|\frac{\langle \{k\} |\Psi_0\rangle}{\langle \{k\} |\Psi_\sigma\rangle}\right| = 1 - \frac{E_k}{\Delta_i}\epp
\end{equation}

The second order term is a bit more involved. The first term in Eq.~\eqref{eq:second_order} can be computed as 
\begin{multline}
  \sum_{\ket{m^{(0)}},\ket{n^{(0)}}\neq \ket{\Psi_{\pm\sigma}}} \ket{m^{(0)}} \frac{\bra{m^{(0)}} H_{\text{xx}} \ket{n^{(0)}}}{E_\sigma - E_m^{(0)}}\frac{\bra{n^{(0)}} H_{\text{xx}} \ket{\Psi_{\sigma}}}{E_\sigma - E_n^{(0)}} 
  \\ =\; - \sum_{\ket{m^{(0)}}\neq \ket{\Psi_{\pm\sigma}}}\sum_{\ket{n^{(0)}}} \ket{m^{(0)}} \frac{\bra{m^{(0)}} H_{\text{xx}} \ket{n^{(0)}}}{E_\sigma - E_m^{(0)}} \frac{\delta_{\ket{n^{(0)}}\in S_1} + \sigma \delta_{\ket{n^{(0)}}\in S_2} }{2\sqrt{2}}\epp
\end{multline}
For each fixed state $\ket{n^{(0)}}\in S_1 \cup S_2$ there are of order $L^2$ many terms for which the flipped spins 
do not connect to each other and for which the energy difference is simply $E_\sigma - E_{m}^{(0)} = -2$. In contrast, 
there are only of order $L$ many terms for which the energy difference is not given by this value. We drop these terms 
because they are subleading in $L$. Since the second and the third term in Eq.~\eqref{eq:second_order} are also 
negligible with respect to the leading contribution, we obtain
\begin{equation}\label{eq:second_order_result}
  \ket{\Psi_0^{(2)}} = - \frac{1}{2} (H_{\textnormal{xx}})^2 \ket{\Psi_\sigma} \epp
\end{equation}
The normalized overlap therefore reads, up to second order in $1/\Delta_i$ and by neglecting the $1/L$ contributions, 
\begin{equation}\label{eq:overlap_ratio_up_to_second_order}
  \left|\frac{\langle \{k\} |\Psi_0\rangle}{\langle \{k\} |\Psi_\sigma\rangle}\right| = 1 - \frac{E_k}{\Delta_i} + \frac{1}{2}\left(\frac{E_k}{\Delta_i}\right)^2 - \ldots \epp
\end{equation}
This provides a second order correction for the nontrivial part of the overlap distribution 
in the quench protocol from $\Delta_i \gg 1$ to $\Delta_f=0$. 

Note that the energy $E_k$ is extensive and bounded by $|E_k| \leq L/\pi$. So, if the condition
$L/\pi \ll \Delta_i$ is fulfilled, we expect the perturbation series to converge. The data for open chains from numerical exact diagonalization, see
Fig.~\ref{fig:fromlarge}, is described reasonably well by the large-$L$ perturbative results derived
in this section for periodic chains.

\section{Conclusions} \label{sec:concl}

Motivated by the observation that the overlap distribution is important for many aspects of quench
dynamics, in this work we have considered a prominent paradigm model, the XXZ chain, and undertaken
a systematic study of the properties of the overlap distribution for quenches of the anisotropy
parameter $\Delta$.  Using numerical exact diagonalization, we have provided a broad overview of the
features of the overlap distribution, for quenches from various values $\Delta_i$ to various values
of $\Delta_f$.

In particular, we have thoroughly analyzed the quench from $\Delta_i=\infty$ to $\Delta_f=0$,
i.e., when the initial state is of the form \eqref{eq:initial} (combination of N\'eel states) and
the post-quench Hamiltonian is described by a free-fermion theory through Jordan-Wigner
transformation.  The overlap distribution in this case has the remarkable feature of being `flat':
all nonzero overlaps have the same value.  Using the free-fermion structure of the post-quench
eigenstates, it was possible to analyze the overlap distribution analytically,
to express the overlaps using determinant formulas, and to explicitly calculate their values (see
Sec.~\ref{sec:B}). The states that have nonzero overlap can be chosen using certain `selection
rules'.  One rule arises from the spatial structure of the N\'eel states, while a second rule arises
from the symmetry of the initial state \eqref{eq:initial}, in terms of translation (periodic
boundary conditions) or reflection (open boundary conditions).  The `selection rules' for the
periodic chain have been compared with results that we derived from recently obtained Bethe Ansatz
formulas \cite{2014_Pozsgay_JStatMech_P06011, 2014_Brockmann_JPA_47_145003,
  2014_Brockmann_JPA_47_345003}.  We found that these rules can be equivalently expressed in terms
of adjacent corners of rectangles in the pattern of possible XX Bethe roots.  We also provided two
example applications of our detailed understanding of the overlap distribution for the
$\Delta_i=\infty\to\Delta_f=0$ quench.  Using the selection rules, we were able to provide concise
expressions for the Loschmidt echo (return probability) as well as equal-time
two-point correlators after the quench.  In particular, time-dependent transverse correlators,
$\langle{}S_{j}^+S_{j+n}^-\rangle$, are somewhat nontrivial due to Jordan-Wigner strings. We provide
determinant formulas for such correlators, derived by making use of the selection rules.

For quenches from  $\Delta_i=\infty$ to small  $\Delta_f>0$, the overlap distribution shows a
splitting structure which we analyzed perturbatively (perturbation in  $\Delta_f$,
see Sec.~\ref{sec:A}).  For quenches from large $\Delta_i\gg1$ to  $\Delta_f=0$, the overlap
distribution acquires a simple dependence on energy which we analyzed using perturbation theory in
$1/\Delta_i$, see Sec.~\ref{sec:C}. 

The present work is related in various ways to recent advances in non-equilibrium physics.  Many
recent publications have provided overlap distributions for specific quenches. This motivates
systematic studies like the present one.  Using the Bethe Ansatz, formulas are now available
\cite{2014_Pozsgay_JStatMech_P06011, 2014_Brockmann_JPA_47_145003, 2014_Brockmann_JPA_47_345003} for
the overlap between N\'eel states and Bethe eingestates at any $\Delta$.  Since the use of these
results requires knowledge of the Bethe roots describing all eigenstates, it is technically challenging to
explore overlap distributions using this approach.  Our numerical overview and detailed explicit
analysis of specific cases thus provides specific contexts for these general technical results.  Our
applications to Loschmidt echoes and correlators can be compared with existing calculations of the
Loschmidt echo \cite{2006_Quan_PRL_96_140604, 2014_Andraschko_PRB_89_125120} and of longitudinal
correlators \cite{2010_Barmettler_NewJPhys_12_055017}).  In contrast to these approaches, we have
used the structure of the overlap distribution to derive time-dependent quantities. This has allowed
us to obtain some new expressions.  Finally, certain overlaps with low energy eigenstates in the
gapless phase $|\Delta_f|<1$ can be studied using field-theoretical methods. In particular,
subleading terms in such overlaps \cite{1991_Affleck_PRL_67_161, 2011_Stephan_PRB_84_195128,
  2012_Bondesan_NuclPhysB_862_553} allow to access universal data about the underlying conformal
field theory.  Our focus here, in contrast, has been on the full spectrum, not just low-energy
physics.

This work opens up many new questions.  First of all, the present study is by no means exhaustive. 
There are features of the overlap distribution in the XXZ chain which remain to be analyzed in
detail.  For example, the physical nature of the eigenstates which have large overlap for
$\Delta_i=0$ to $\Delta_f$ of order one quenches (or vice versa), are
unclear and might deserve further study.  Second, in addition to the applications we have provided
for the cleanest case ($\Delta_i=\infty\to\Delta_f=0$). It might be possible to exploit the
observations on overlap distributions in other quench protocols (other $\Delta_i$ and $\Delta_f$) to
extract properties of the time dependence of Loschmidt echoes and correlation functions.  Finally,
it would certainly be interesting to analyze overlap distributions in a series of other models.
One could ask, for example, whether integrability plays a general role in determining the features
of the overlap distribution.

\section*{Acknowledgements}
We thank Jacopo Viti for useful discussions. 
V.A.~acknowledges support by the ERC under the Starting Grant 279391 EDEQS.


\end{document}